\documentclass[reprint,superscriptaddress, prl, aps]{revtex4-2}
\usepackage{graphicx}
\usepackage{tikz}
\usepackage{dcolumn}
\usepackage{times}
\usepackage[braket, qm]{qcircuit}
\usepackage{physics}
\usepackage{mathtools}
\usepackage{wasysym}
\usepackage{amsmath}
\usepackage{amssymb}
\usepackage{hhline}
\usepackage{multirow}
\usepackage{hhline}
\usepackage{xcolor}
\usepackage[final]{listings}
\usepackage{hyperref}
\usepackage{graphicx}

\definecolor{darkgreen}{rgb}{0,0.5,0}
\definecolor{darkred}{rgb}{0.5,0,0}
\definecolor{darkblue}{rgb}{0,0,0.5}
\definecolor{MediumGray}{gray}{0.60}

\lstdefinestyle{pseudo}{
	language=Python,
	xleftmargin=5.0ex,
	moredelim = [s][\textit]{[}{]},
	basicstyle={\small\ttfamily},
	captionpos=b,
	columns=flexible,
	numbers=left,
	numberstyle=\small,
	numbersep=8pt,
	stepnumber=1,
	numberstyle=\tiny\color{gray},
	escapechar=\%,
	breaklines=true,
	frame=single,
	framexleftmargin=15pt,
	tabsize=2,
	postbreak=\mbox{\textcolor{red}{$\hookrightarrow$}\space},
	captionpos=b,
	showspaces=false,
	showtabs=false,
	keywords=[3]{QuantumState, List, Tuple},
	commentstyle=\itshape\color{MediumGray},
}

\begin{document}

\title{Realization of real-time fault-tolerant quantum error correction}

\author{C. Ryan-Anderson, J. G. Bohnet, K. Lee, D. Gresh, A. Hankin, J. P. Gaebler, D. Francois, A. Chernoguzov, D. Lucchetti, N. C. Brown, T. M. Gatterman, S. K. Halit, K. Gilmore, J. Gerber, B. Neyenhuis, D. Hayes, and R. P. Stutz}
\affiliation{Honeywell Quantum Solutions, Broomfield, CO }

\date{\today}

\begin{abstract}
Correcting errors in real time is essential for reliable large-scale quantum computations.
Realizing this high-level function requires a system capable of several low-level primitives, including single-qubit and two-qubit operations, mid-circuit measurements of subsets of qubits, real-time processing of measurement outcomes, and the ability to condition subsequent gate operations on those measurements.
In this work, we use a ten qubit QCCD trapped-ion quantum computer to encode a single logical qubit using the $[[7,1,3]]$ color code, first proposed by Steane~\cite{steane1996error}.
The logical qubit is initialized into the eigenstates of three mutually unbiased bases using an encoding circuit, and we measure an average logical SPAM error of $1.7(2) \times 10^{-3}$, compared to the average physical SPAM error $2.4(8) \times 10^{-3}$ of our qubits.
We then perform multiple syndrome measurements on the encoded qubit, using a real-time decoder to determine any necessary corrections that are done either as software updates to the Pauli frame or as physically applied gates.
Moreover, these procedures are done repeatedly while maintaining coherence, demonstrating a dynamically protected logical qubit memory.
Additionally, we demonstrate non-Clifford qubit operations by encoding a $\overline{T}\ket{+}_L$ magic state with an error rate below the threshold required for magic state distillation.
Finally, we present system-level simulations that allow us to identify key hardware upgrades that may enable the system to reach the pseudo-threshold.
\end{abstract}
\maketitle

\section{Introduction}
Large-scale quantum computers promise to solve classically intractable problems in areas such as quantum simulation, prime factorization, and others \cite{feynman1986quantum, shor1999polynomial, shor1994algorithms, abrams1997simulation, aspuru2005simulated, von2020quantum, orus2019quantum}. However, these complex quantum computations demand levels of precision that are currently unachievable due to imperfect control and noise in gate operations between physical qubits. In fact, it
is unlikely that analog physical qubit control will ever be able to reach the precision demanded by large scale computations. Quantum error correction (QEC) ~\cite{shor1995scheme, calderbank1996good, steane1996error}, was the key ingredient to digitize quantum operations, making extremely low error rates possible in principle. QEC works by redundantly encoding quantum information into a protected subspace within a larger Hilbert space of many physical qubits. Using a polynomially scaling number of physical qubits, the probability of a computation being corrupted can be suppressed exponentially, making arbitrarily precise quantum computation feasible.

Achieving efficient error suppression by a QEC code introduces several requirements for a quantum processor. Not only do the error rates of the underlying physical operations (initialization, unitary gates, and measurement) need to be below a certain threshold \cite{aharonov2008fault, kitaev1997quantum, knill1996threshold}, but the quantum processor must interact with a classical computer in real time to diagnose and correct errors. These interactions between the quantum and classical processors need to be repeated several times in every step of a computation, defining new requirements in addition to the classic DiVincenzo criteria \cite{divincenzo2000physical}.
These new criteria include the ability to measure a subset of qubits with little impact on other qubits and the ability to classically process measurements and determine corrections faster than the system decoheres.
Additionally, the implementation should be fault-tolerant (FT) to at least the dominant physical layer errors in order to prevent physical errors from cascading and causing logical errors~\cite{gottesman1997stabilizer, gottesman2010introduction, gottesman1998theory}.

Several required elements of FT QEC of a single logical qubit have been demonstrated on a variety of quantum computing architectures: classical repetition codes \cite{chen2021exponential, cory1998experimental, chiaverini2004realization, schindler2011experimental, cramer2016repeated, reed2012realization, riste2015detecting, riste2020real, wootton2018repetition}, error detection codes \cite{Linkee1701074, PhysRevLett.119.180501, corcoles2015demonstration, harper2019fault, kelly2015state, andersen2020repeated, erhard2020entangling}, the 5-qubit code \cite{gong2021experimental, knill2001benchmarking}, the [[7,1,3]] color code \cite{nigg2014quantum, hilder2021faulttolerant}, the Bacon-Shor-13 code \cite{egan2020fault},
the 9-qubit Shor code \cite{luo2020quantum, aoki2009quantum}, Bosonic codes \cite{hu2019quantum, heeres2017implementing, fluhmann2019encoding, ofek2016extending, campagne2020quantum}, as well as primitives utilizing cluster states \cite{yao2012experimental, lu2008experimental}.
However, a full demonstration of all necessary components for a FT implementation of a QEC code capable of repeatedly correcting all single-qubit errors has not yet been realized.

In this article, we report on realizing a FT implementation of the smallest instance of the color code~\cite{steane1996error,bombin2006topological,reichardt2020fault,Bermudez_2019}. Using a trapped-ion QCCD quantum computer~\cite{Kielpinski02, Pino2020} we encode, control, and repeatedly error correct a single logical qubit using 10 physical qubits. On the physical layer, trapped-ion qubits use high-fidelity single and two-qubit gates and mid-circuit measurements and resets to execute the quantum circuits that fault-tolerantly initialize the logical qubit, manipulate it via logical single-qubit Clifford gates, perform error syndrome measurements, and apply corrections. Importantly, the low crosstalk during mid-circuit measurements~\cite{chiaverini2004realization,Barrett04} enable ancilla measurements of the error syndromes without decohering the data qubits that encode the logical information. The error syndrome measurements are sent to a classical computer where real-time decoding and tracking of errors and corrections takes place. Since all of this can be done with high-fidelity and quickly compared to the dephasing rate of the physical qubits, the logical qubit can be \textit{repeatedly} error corrected, a crucial feature of scalable quantum computing. We perform these operations with the six eigenstates of the logical Pauli operators and intialize a magic state for non-Clifford operations. These results demonstrate a universal set for quantum computation, with the notable exception of an entangling gate between two logical qubits, which requires more qubits than our system currently supports.

\subsection{Background}
\begin{figure}[h]
\includegraphics[trim=0 5 190 135, clip,width=\columnwidth]{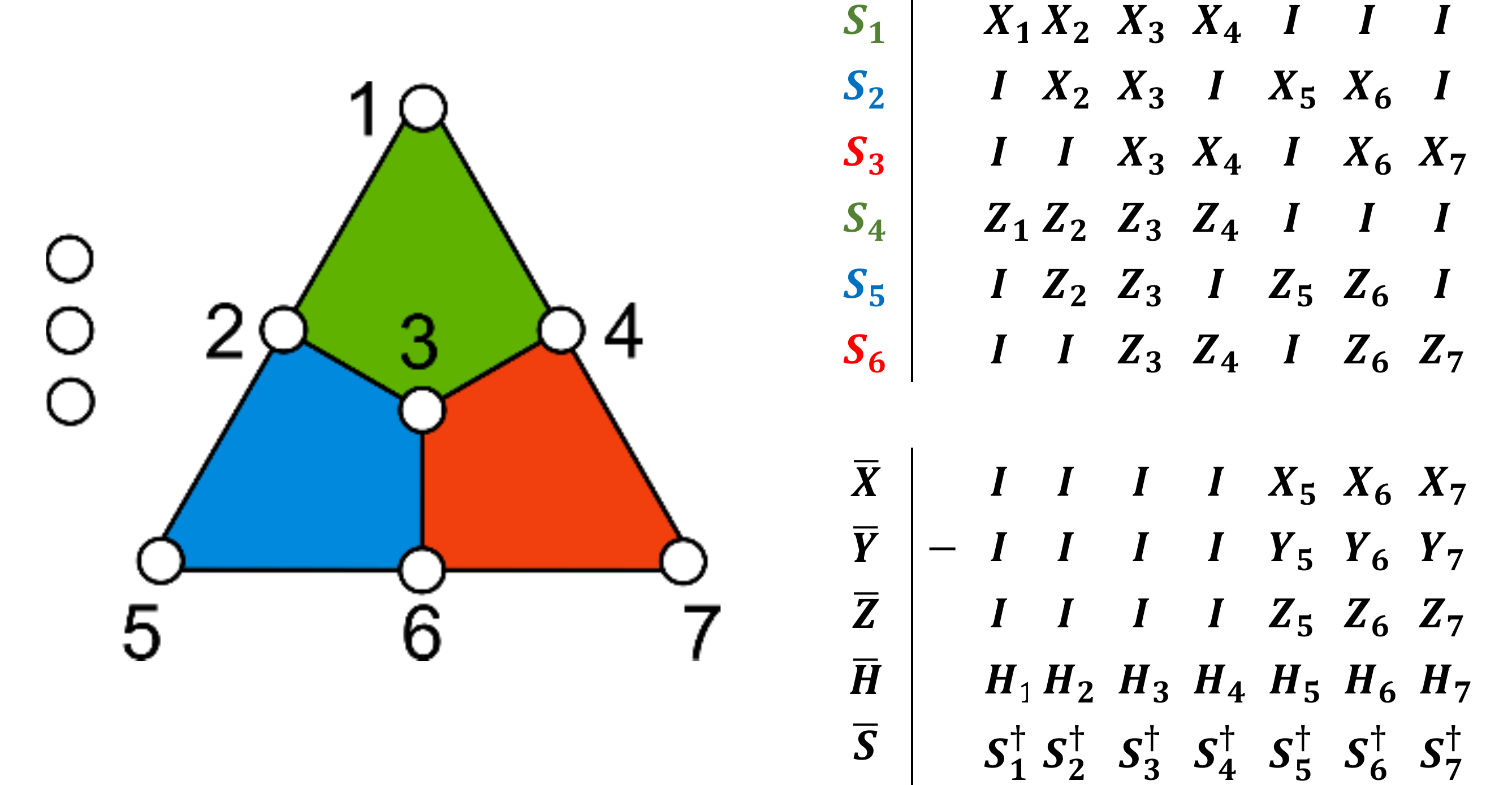}
\caption{The $[[7,1,3]]$ color code. The seven data qubits are on the vertices of the polygons and three ancilla qubits for syndrome measurements are off to the side. For each polygon, the four qubits at the vertices define both $X$-type and $Z$-type stabilizer measurements used in each error correction cycle. The stabilizer generators and logical operators are tensor products of single qubit Paulis with support indicated by the physical qubit subscript index. For example, implementing the logical $\overline{Z}$ operation is done with physical layer $Z$ operations on qubits 5, 6, and 7. Likewise, measuring $\overline{Z}$ is done by measuring $Z_5Z_6Z_7$ where the logical qubit measurement outcome is the product of the three individual physical qubit measurement outcomes. Although not shown, $Y$-type stabilizers are generated by the $X$ and $Z$-type stabilizers listed here.}
\label{steane_code}
\end{figure}

QEC codes are identified by three parameters $[[n,k,d]]$, where $n$ is the number of physical qubits, $k$ is the number of logical qubits the code admits, and $d$ is the code distance, which is related to the minimum number of arbitrary single-qubit errors the code will correct $t=\lfloor\frac{d-1}{2}\rfloor$. The smallest example of the topological color code \cite{bombin2006topological}, depicted in Fig.~\ref{steane_code}, is a $[[7,1,3]]$ stabilizer code, sometimes referred to as the Steane code. The distance three code uses seven physical qubits to encode a single logical qubit and protects against all instances of a single physical qubit incurring an error ($t=1$). Advantages of topological stabilizer codes include: requiring only local interactions, high error thresholds, and minimal error detection overhead~\cite{dennis2002topological, bombin2006topological, fowler2012surface}. The $[[7,1,3]]$ color code has the added advantage of all single-qubit and two-qubit Clifford gates being transversal, which are naturally FT~\cite{Nielsen00}. While the term ``color code'' generally refers to any member of the family of color codes, we only investigate one version here, so for simplicity we drop the $[[7, 1, 3]]$ notation and refer to the code we study as the color code. We also note that we place overbars and $L$ subscripts on logical qubit operators and states to distinguish them from physical qubit operators and states.

The color code, like all stabilizer codes, detects errors by measuring a commuting set of operators known as stabilizers (Fig.~\ref{steane_code}).
The stabilizer measurements form an error syndrome which is processed by a classical decoding algorithm to determine a correction, a process that must be done in real-time for any non-trivial logical computation. We henceforth define a QEC cycle as the process of measuring syndromes (including measuring multiple times to account for measurement errors), decoding, and applying corrections.

\begin{figure}[h]
\includegraphics[width=\columnwidth]{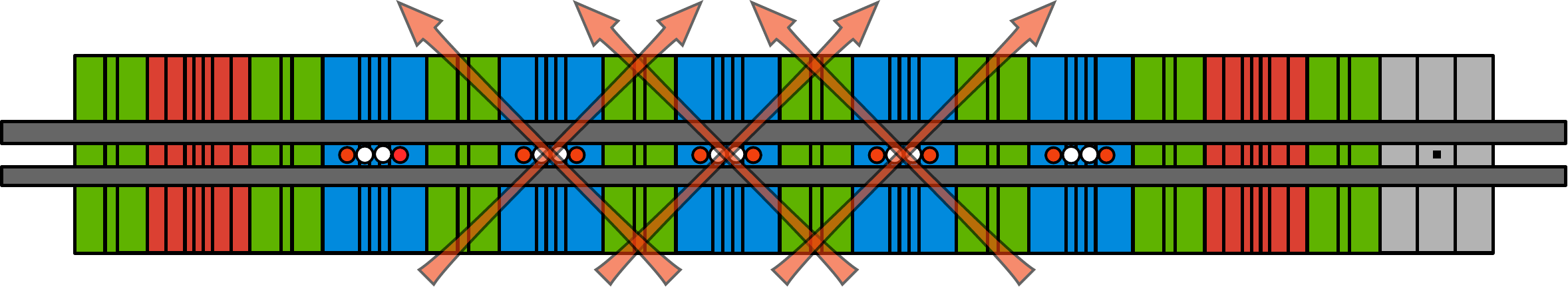}
\caption{The ion trap loaded with ten $^{171}$Yb$^+$ qubit ions (red circles) and ten $^{138}$Ba$^+$ coolant ions (white circles). The trap has different functional regions, or zones, with functions determined by the electrode geometry and laser beam configuration. Ion transport is used to arrange ions into zones for gates and measurements. The electrodes in red and blue denote regions that support transport operations including linear transport, crystal split and combine, and physical swap. The green regions support linear transport and storage of ions between gating operations. The three regions where qubit initialization, gating, and measurement occur are marked by the crossing laser beams. In the grey region, ions are loaded from an effusive atomic oven behind the trap using photoionization.}
\label{fig:Trap}
\end{figure}

Our quantum computer uses a Honeywell 2D surface electrode ion trap (Fig.~\ref{fig:Trap}) similar to the one described in Ref.~\cite{Pino2020}.
The trap is loaded with ten $^{171}$Yb$^+$ qubit ions and ten $^{138}$Ba$^+$ sympathetic cooling ions. The color code is implemented using all ten available qubits, (Fig. \ref{steane_code}), where seven data qubits encode the logical state and three ancilla qubits perfom syndrome measurements.
The trap can execute parallel quantum operations across three different zones and quantum circuits are executed as a series of interleaved initialization, gating, measurement, and ion-transport operations.
With just three ion-transport primitive operations, qubits can be arbitrarily rearranged in the middle of a circuit: qubit ions paired with coolant ions use (1) linear transport of $\{\mathrm{Yb,Ba}\}$ and $\{\mathrm{Yb,Ba,Ba,Yb}\}$ crystals, (2) split/combine operations to go between $\{\mathrm{Yb,Ba}\}$ crystals and $\{\mathrm{Yb,Ba,Ba,Yb}\}$ crystals, and (3) physical (not quantum gate-based) swap operations to switch the ordering of the qubit ions in a crystal by transporting them around each other in two dimensions~\cite{Splatt_2009}. These operations allow any two qubits to be paired for entangling gates, and allow qubits to be isolated for single-qubit gates, initialization and measurement.
Dynamically rearranging the ions during the circuits ensures the one-dimensional geometry trap does not restrict the geometry of codes that can be used (the color code has a two-dimensional geometry).

The system is programmed according to the quantum circuit model.
Logical qubit operations are compactly described at the physical qubit level as quantum circuits and are expressed in an extended version of OpenQASM 2.0~\cite{cross2017open}.
At the time of developing the QEC experiments discussed in this paper, the OpenQASM language did not fully support all of the conditional logical operations needed for QEC. We therefore extended the language to include classical assignment, classical operations, and expanded comparison of registers. These OpenQASM extensions were key enablers for the highly dynamical QEC protocols we chose to implement \cite{reichardt2020fault}.

\section{Experiments}

In this section we discuss our experiments demonstrating the QEC operations necessary for universal QEC computation restricted to a single logical qubit.

\subsection{QEC cycles}
The main result of this paper is a full demonstration of the ability to repeat QEC cycles, which includes the determination of corrections during computation. This demonstration consists of initializing a logical Pauli basis state using a FT encoding circuit, running multiple adaptive FT QEC cycles, and then measuring in the appropriate logical basis, all while tracking corrections, running a decoding algorithm, and updating the corrections after each QEC cycle in real-time. A schematic of the overall procedure can be see in Fig. \ref{FlowChart}. We now detail the various steps of this QEC protocol.
\subsubsection{Logical state preparation}
To prepare different logical basis states, $\{\ket{0}_L,\ket{1}_L,\ket{+}_L, \ket{-}_L,\ket{+i}_L,\ket{-i}_L\}$, we utilize a FT encoding circuit. The encoding circuit first prepares $\ket{0}_L$ and fault-tolerantly verifies successful preparation by measuring $\overline{Z}$~\cite{goto2016minimizing} via three CNOTS with an ancilla at the end as shown in Fig.~\ref{FlowChart}.
If the measured ancilla is found in $\ket{0}$, the circuit succeeded and moves to the next step. If the ancilla is found in $\ket{1}$, all qubits are reinitialized and the circuit runs again. This procedure repeats until successful, up to three iterations if necessary.
This verification step succeeds with a probability of approximately $99.984(9) \%$ with the first attempt succeeding approximately $97.9(2) \% $ of the time. Regardless of successful preparation, we proceed by rotating $\ket{0}_L$ to another logical Pauli basis-state by the appropriate application of the $\overline{X}$, $\overline{H}$, and $\overline{S}$ operators as prescribed in Fig. \ref{steane_code}.

\subsubsection{Adaptive syndrome extraction protocol}
Next, to protect the logical qubit while idling during a computation, we execute multiple QEC cycles and utilize ancilla qubits to measure syndromes in a non-destructive manner.
The syndrome extraction protocol operates by detecting changes in stabilizer measurement outcomes that ideally give a +1 eigenvalue.
Because syndrome extraction requires the use of noisy gates, syndrome measurements must be repeated multiple times within a single QEC cycle to be FT to syndrome measurement errors. For each FT QEC cycle, we implement the ‘flagged’ three-parallel
syndrome extraction protocol described in Ref.~\cite{reichardt2020fault} (see Fig. \ref{FlowChart} and Pseudocode \ref{code:qec-cycle_exp}).

 In order for the color code to be FT to all single qubit errors, we must account for a special case of a single qubit error, called a `hook' error \cite{dennis2002topological}, which spreads to higher weight errors causing logical errors, see Fig. \ref{hook_decoder} \textit{a}.
The syndrome extraction protocol measures stabilizers in a way that flags (identifies) malignant `hook' errors.
The protocol first uses flagged parallel circuits to measure syndromes, the first set being $\{S^f_1, S^f_5, S^f_6\}$.
If there are no changes in stabilizers measurements (\textit{i.e.}, no errors indicated), the second set of flagged stabilizers $\{S^f_1,  S^f_3,  S^f_4\}$ is also measured, and again, if there are no changes in stabilizers measurements, the syndrome extraction protocol is deemed complete.
However, if either of the flagged circuits do indicate an error, an additional and final round of syndrome extraction is triggered using standard circuits without flags  $\{S_1,S_2,S_3, ... \}$ (Fig.~\ref{FlowChart}).
Note that both sets of stabilizers follow from the definitions given in Fig. \ref{steane_code}, but we use the $f$ superscript to differentiate between syndromes measured using flagged circuits and unflagged circuits.
The exact circuits used along with pseudocode can be found in the supplemental material.
During every QEC cycle, syndrome changes are sent to a decoder to infer a correction and the baseline stabilizer values are updated in software for the next QEC cycle.

\subsubsection{Decoder}
Our decoder algorithm consists of two steps using two different lookup tables, analogous to the decoders described in~\cite{chao2018quantum, chao2018fault}.
Lookup tables are simple decoders that map syndromes to corrections.
We decode $X$ and $Z$ errors separately and correct at the logical level instead of at the physical level (that is, the corrections are whether to apply $\overline{I}$, $\overline{X}$, $\overline{Y}$, or $\overline{Z}$). These choices reduce the size of the decoder to a few if-statements in QASM
(see Pseudocodes~\ref{code:decoder-2d} - \ref{code:decoder-fix}).
During a particular QEC cycle, the decoder is called if and only if the final round of syndrome extraction is triggered (\textit{i.e.}, the unflagged circuits, see Fig. \ref{FlowChart}).

The first stage in the decoding algorithm uses changes in the unflagged syndrome measurements $S_i$ \textit{between} QEC cycles; specifically, the final round of syndrome extraction in the current QEC cycle versus the last QEC cycle that triggered the execution of the unflagged circuits.
It is only necessary to consider changes in syndrome measurements between QEC cycles because after an error changes the syndrome measurement, the subsequent corrections do not change the syndromes back, as syndrome operators commute with correction operators.
Note, the last QEC cycle that triggered a final round of syndrome extraction may or may not be the proceeding cycle.
If no previous rounds of a QEC cycle triggered the unflagged checks, the syndromes is assumed to be trivial.
This lookup table considers spatial but not temporal errors to produce the initial logical correction.

 The second stage of the decoding algorithm uses changes in the syndromes between extraction rounds \textit{within} the current cycle.
 Single qubit hook errors produce the same syndrome measurements in the unflagged circuits as other less damaging single qubit errors (see Fig. \ref{hook_decoder}).
 Thus, to distinguish between these two types of single qubit errors, we require additional syndrome information provided by the flag circuits.
 We compare the two sets of syndrome information ($S^f_i$ vs $S_i$) and use a second lookup table to determine the final logical correction.
The combination of the conditional rounds of syndrome extraction and the second lookup table makes the QEC cycle FT to both hook and measurement errors.

\subsubsection{Pauli frame update}
Applying corrections physically via noisy gates can potentially induce more errors.
Fortunately, many errors can be corrected without implementing physical gate operations and instead done with an essentially perfect software correction.
To this end, the correction for the logical qubit is stored in a binary array, known as a Pauli frame \cite{knill2005quantum}, during the computation.
The Pauli frame is represented by two bits, corresponding to the possible corrections $\{\overline{I},\overline{X},\overline{Y},\overline{Z}\}$.
At the end of each QEC cycle, the new correction is combined with the previous Pauli frame according to Pauli matrix multiplication rules, and the binary array is updated.

Finally, after completing a variable number of QEC cycles, the data qubits are directly measured. The final destructive measurement projects the state into the logical $\overline{X}$, $\overline{Y}$, or $\overline{Z}$ basis based on the logical single-qubit gate applied before measurement, and from this measurement two pieces of critical information are extracted. First, the raw logical output is calculated by multiplying the $\pm1$ outcomes of the non-identity operations of the logical operator being measured (qubits 5, 6, and 7, see Fig.\ref{steane_code}). Second, while the process of measuring the data qubits is destructive to the quantum state, the syndromes can be constructed from the resulting classical information, allowing for a final correction.
For example, when measuring the fidelity of $\ket{+}_L$, we measure all data qubits in the $X$ basis and the raw logical output is given by results of qubits 5, 6, and 7 while the syndrome is equal to the measurement result tuple of $\{S_1,S_2,S_3\}$.
However, while for $\overline{X}$ and $\overline{Z}$ measurements it is sufficient to infer the syndomes using stabilizers of the same Pauli type in Fig.~\ref{steane_code}, this is not the case for the $Y$ basis. Fortunately, in the special case of a $\overline{Y}$ measurement, we can infer the syndromes of the three $Y$ stabilizers that are the products of the $X$ and $Z$ stabilizers listed in Fig.~\ref{steane_code} that have identical non-trivial support.
After the syndromes are inferred from the final measurement, the decoding algorithm is used to determine a correction, simply whether to flip the raw logical output or not.

\subsubsection{Results}

\begin{figure}
	\includegraphics[width=\columnwidth]{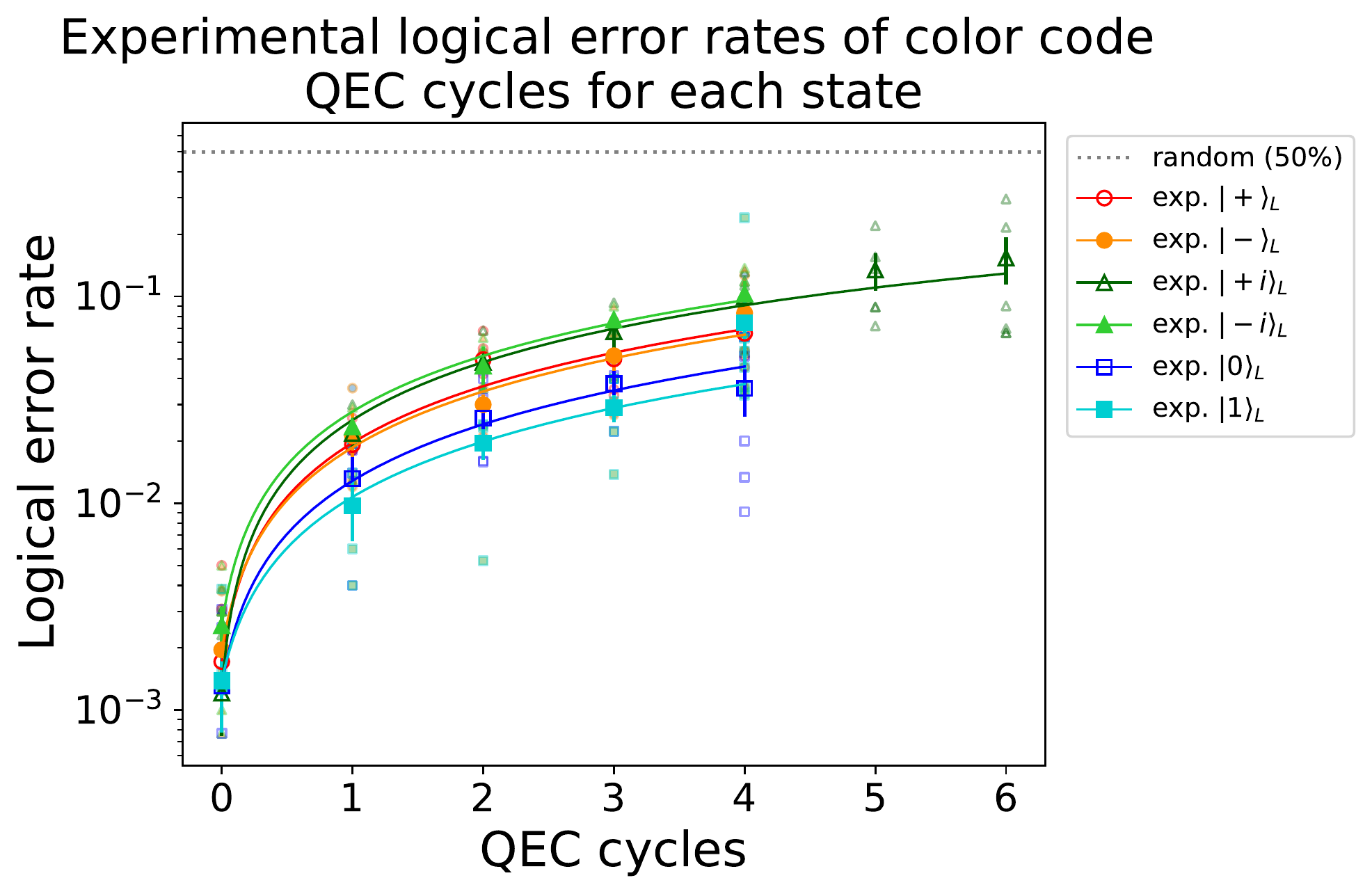}
	\caption{Comparing the observed logical fidelities of the six logical Pauli basis states over many QEC cycles. Averages and standard deviations were determined by Jackknife resampling between individual experiments \cite{efron1982jackknife}. The large points with error bars are experimental averages and standard deviations. The smaller and lighter points are individual trials for a given QEC cycle experiment. Note, for 0 QEC, numerous experimental trials have an error rate of 0 and, thus, are not displayed due to the log scale. The lines are fits to the experimental averages, where the fits are exponential decay curves $p_L(c)=0.5+(p_{spam}-0.5)\left(1-2 \; p_{cycle}\right)^c$. Here $p_{L}(c)$ is the logical error rate of a cycle $c$ and logical basis $L$, $p_{spam}$ the logical SPAM error, and $p_{cycle}$ is the logical QEC cycle error. Note, $p_{spam}$ is determined directly from the 0 QEC results and fixed when determining the fits. Thus, from the fits we obtain estimates of $p_{cycle}$. The values of $p_{spam}$ and $p_{cycle}$ are reported in Table~\ref{SPAM_fid_exp}.
	}
\label{log_err_exp}
\end{figure}

The experimental results for the full QEC protocol can be seen in Fig. \ref{log_err_exp}.
Each basis state and QEC cycle iteration was run multiple times with a varying number of trials.
To better estimate the error bars due experimental noise fluctuations, we used a Jackknife resampling method \cite{efron1982jackknife} to estimate the average and the standard deviation of the different trials for a given state.
To estimate the logical SPAM and logical error rate per QEC cycle, we fit the data to a exponential decay curve and extract the fitting parameters as shown in Table \ref{SPAM_fid_exp}.
In the protocol, logical SPAM is equivalent to doing zero QEC cycles.
We measured the average logical SPAM error to be $1.7(6) \times 10^{-3}$, compared to the $2.4(8) \times 10^{-3}$  SPAM error of our physical qubits.
The data in Fig.~\ref{log_err_exp} shows that while the initialization circuit produces high-fidelity states, repeated QEC cycles introduce significant logical errors of approximately $2.70(6)\%$ per QEC cycle as determined by an exponential decay fit to the data. We observe that the logical $\overline{Z}$ basis is more robust than the other two bases, suggesting that qubit dephasing is a significant error source.

The experimental run time of the logical SPAM portion of experiments, including the encoding circuit, two rounds of single-qubit rotations, and the final measurement, is $<60$ ms, and each QEC cycle takes less than $<200$ ms. Transport accounts for $\sim10$ ms of logical SPAM operations and $<70$ ms of each QEC cycle. The remaining time is dominated by the cooling operations, which occur prior to each two-qubit gate, similar to the experiments detailed in Ref.~\cite{Pino2020}.

To put the logical qubit error rates and clock-speed into context by comparing to the physical layer; the dominant error at the physical level is the two-qubit gate error of $\sim3\times10^{-3}$ (Table \ref{Sim_Parameters}), and the physical layer clock-speed (one round of gating between random pairs of qubits) is $10.5$ ms. We note that a complete understanding of the overhead associated QEC will have to wait until logical qubit entangling operations are characterized.

\begin{table}[]
\begin{tabular}{|l|| c| c|}
\hline
 \textbf{State} & \textbf{$p_{spam}$} &\textbf{$p_{cycle}$}\\
 \hline
 $\ket{+}_L$& $1.7(7)\times10^{-3}$& $ 1.8(1)\times10^{-2}$\\
 \hline
 $\ket{-}_L$& $ 2.0(7)\times10^{-3}$& $ 1.7(2)\times10^{-2}$\\
 \hline
 $\ket{+i}_L$& $1.2(5)\times10^{-3}$& $ 2.4(1)\times10^{-2}$\\
 \hline
 $\ket{-i}_L$& $2.6(6)\times10^{-3}$& $ 2.5(1)\times10^{-2}$ \\
 \hline
 $ \ket{0}_L$ & $1.3(5)\times10^{-3}$ & $1.16(7)\times10^{-2}$\\
 \hline
 $\ket{1}_L$& $1.4(6)\times10^{-3}$& $ 0.94(7)\times10^{-2}$\\
 \hline
\end{tabular}
\caption{Observed logical SPAM error $p_{spam}$ from 0 QEC experimental results. And logical error per QEC cycle $p_{cycle}$ fit parameters for the six Pauli basis states, obtained via exponential decay fit and used in the plotted curves in Fig.~\ref{log_err_exp}. Where the average $p_{spam}$ is  $1.7(2)\times10^{-3}$ and the average $p_{cycle}$ is $1.75(4)\times10^{-2}$]}
\label{SPAM_fid_exp}
\end{table}

\subsection{Active vs software corrections}
 Pauli frame updates cannot always be used for a QEC computation since Pauli operators have non-trivial transformations under conjugation by non-Clifford gates. Thus, to implement corrections stored in software, we either physically apply the correction to the qubits before the gate, or adapt the non-Clifford gate to include the correction.
To demonstrate the ability to physically apply corrections as needed, we use the logical S gate as a stand in for the non-Clifford T gate. We first initialized the state $\ket{+}_L$, ran one cycle of QEC to generate potential corrections, physically applied a correction (either $\overline{X}$, $\overline{Y}$, or $\overline{Z}$) according to the Pauli frame, physically applied logical $\overline{S}$, performed an additional QEC cycle and measured in the $\overline{Y}$ basis (note $\overline{S}\ket{+}_L = \ket{+i}_L$), see Fig. \ref{S_gate}), achieving a logical fidelity of $93(2)\%$.

We also repeated this experiment without applying a physical correction and instead rotated the Pauli frame according to Pauli transformations of the physically applied $S$ gate. This software-correction version of the experiment achieved a logical fidelity of $92(1)\%$. These two error rates are not significantly different from each other, thus demonstrating the key ability to take software tracked corrections and apply them in real-time as necessary.

\subsection{Preparing a magic state}
Universal quantum computing requires the ability to implement non-Clifford gates which, unlike Clifford gates, can not be constructed by simple transversal operations in the color code.
Non-Clifford gates can, however, be implemented using state preparation primitives that produce so-called magic states \cite{bravyi2005universal}.
One choice of magic state that is sufficient to complete a universal gate set is $\overline{T}\ket{+}_L=\left(\ket{0}_L + e^{i \pi /4} \ket{1}_L\right) / \sqrt{2}$, where $\overline{T} = \text{diag}(1,e^{i\pi/4})$.
Note that we cannot use the FT encoding circuit used for the logical Pauli states, because the verification step requires the measurement of a logical operator which would collapse the $T$-state, and thus, we use a non-fault tolerant encoding circuit for the color code \cite{preskill1998reliable}, shown in Fig. \ref{TState}.
Once prepared, $\overline{T}\ket{+}_L$ can be used to apply $\overline{T}$ gates via gate teleportation~\cite{zhou2000methodology} in a system capable of logical two-qubit gates.
The fidelity of this operation is estimated by measuring the operator $\overline{T}\ket{+}_L\bra{+}_L\overline{T}^{\dagger}=\frac{1}{2}\left(\overline{I}+\frac{1}{\sqrt{2}}(\overline{X}+\overline{Y})\right)$ and we report an error of $2.2(6) \%$, significantly lower than the noise threshold estimate of $33.5\%$ for magic state distillation \cite{bravyi2005universal}.
Thus, even with a non-FT encoding circuit, we are able to produce high quality, distillable states to implement FT non-Clifford gates.
\section{Simulations and Analysis}

\begin{figure}
	\includegraphics[width=\columnwidth]{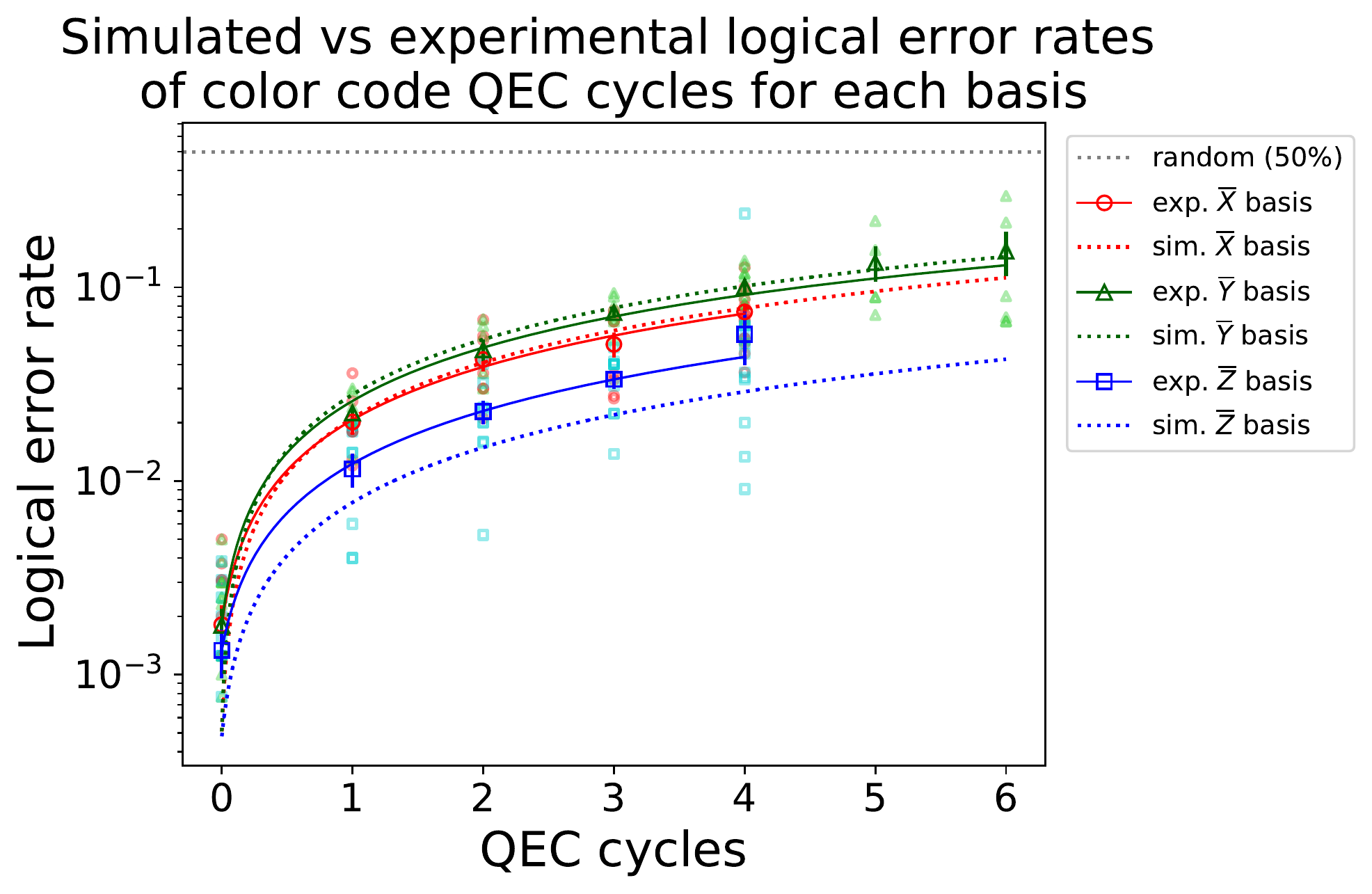}
	\caption{Comparing the observed and simulated logical fidelities of the logical Pauli bases. Averages and standard deviations were determined by combining the data of both basis states of a given basis and using Jackknife resampling between individual experiments \cite{efron1982jackknife}. The large points with error bars are experimental averages and standard deviations. The smaller and lighter points are individual trials for a given QEC cycle experiment. Note, for 0 QEC, numerous experimental trials have an error rate of 0 and, thus, are not displayed due to the log scale. The lines are fits to the simulated and experimental averages, where the fits are exponential decay curves $p_L(c)=0.5+(p_{spam}-0.5)\left(1-2 \; p_{cycle}\right)^c$. Here $p_{L}(c)$ is the logical error rate of a cycle $c$ and logical basis $L$, $p_{spam}$ the logical SPAM error, and $p_{cycle}$ is the logical QEC cycle error. Note, $p_{spam}$ is determined directly from the 0 QEC experiments and fixed when determining the fits. Thus, from the fits we obtain estimates of $p_{cycle}$. The values of $p_{spam}$ and $p_{cycle}$ are reported in Table~\ref{SPAM_fid_sim}.
	}
	\label{log_err_basis}
\end{figure}

To help understand the noise in our system, we compare the experimental results of the QEC protocol to numerical simulations as seen in Fig.~\ref{log_err_basis}. The simulations are state vector simulations~\cite{crathesis,pecos} utilizing a realistic error model and experimentally measured error parameters (for further details see Table \ref{Sim_Parameters}).
The error model includes simple depolarizing gate noise, leakage errors, and modeling of coherent dephasing noise during transport and cooling operations.
The simulation uses the same instructions that are generated by the compiler and sent to the quantum computer, including the same gate decomposition, gate duration, and transport.

\subsection{Comparison to simulations}
The complex physical layer error model can be distilled to a much simpler logical error model described by an asymmetric depolarizing channel,
\begin{equation}
	\overline{\mathcal{E}}(\rho) = (1-p_L) \rho + p_{x} \overline{X} \rho \overline{X} + p_{y} \overline{Y} \rho \overline{Y} + p_{z} \overline{Z} \rho \overline{Z}, \\
	\label{err_chan_eq}
\end{equation}
where the Pauli error probabilities $p_x$, $p_y$, and $p_z$ are fitting parameters for both the experimental and simulation data. Note that $p_L=p_x+p_y+p_z$ ensures the map is trace preserving and $p_L$ is referred to as the logical error rate.

For individual basis states, only the Pauli operators that do not commute with the state cause errors.
For example, the $\overline{Z}$ basis is immune to $\overline{Z}$ type noise but susceptible to $\overline{X}$ and $\overline{Y}$ type noise.
Therefore, the error probabilities associated with the different bases are given by the three equations,

\begin{equation}
\begin{split}
\textbf{$\overline{X}$} \; basis:  p_{yz} = p_y + p_z \\
\textbf{$\overline{Y}$} \; basis:  p_{xz} = p_x + p_z \\
\textbf{$\overline{Z}$} \; basis:  p_{xy} = p_x + p_y.
\end{split}
\label{error_rate}
\end{equation}

By fitting the experimental and simulated data of individual basis states to exponential decay curves, the probabilities $p_{yz}$, $p_{xz}$, and $p_{xy}$ are solved for and are reported in Table~\ref{SPAM_fid_sim}. The system of equations in Eq.~\ref{error_rate} can then be inverted to solve for the depolarizing parameters $p_x$, $p_y$, and $p_z$ and are reported in Table~\ref{err_chan}.

For most of the error parameters we have experimental estimates (see supplemental materials).
However, accounting for different sources of dephasing noise is not straightforward.
To account for dephasing, we vary the dephasing rate used in the simulation and compare with the experimental results. The depolarizing parameters determine the logical error rate $p_L$, which is what we use to empirically determine the best fit dephasing rate.

The simulation and experimental results qualitatively agree, suggesting that the most important sources of noise are understood.
However, further investigation is needed to fully understand the impact of additional known errors, as well as unknown error sources. In particular, it is important that we further characterize both coherent and incoherent phase noise sources independent of QEC. We also note the bias in the logical error model toward the $\overline{Z}$ component and the interesting asymmetry between the $\overline{X}$ and $\overline{Y}$ components. While the $\overline{Z}$ bias may be explained by the asymmetry in the microscopic noise model from the presence of dephasing, there is no such asymmetry between $X$ and $Y$ in the physical layer error model. Therefore, we suspect that the circuit structure tends to convert $Z$ noise in an asymmetric manner, however we leave a detailed analysis to future work.

Since the dephasing errors manifest from different sources (spatial and temporal magnetic field fluctuations, relative phase drifts between different laser beams, etc.) there are likely both coherent and incoherent sources of qubit dephasing in our system.
To gain understanding of the impact of coherent dephasing on the QEC protocol, we ran both coherent (state-vector) and incoherent (stabilizer) simulations.
The coherent dephasing is modeled as the channel
\begin{equation}
R_Z(\theta) \; \rho \; R_Z(\theta)^\dagger = \exp(-i Z \theta/2) \; \rho \; \exp(+i Z \theta/2),
\label{dephasing_chan_eqn}
\end{equation}

which is applied between ideal gates where  $\theta = dephasing \; rate \times duration$. The duration corresponds to the time it takes for transport operations or the qubit idling time while operations are being applied to other qubits. For the incoherent simulation, dephasing is modeled as the Pauli twirled version of the coherent channel, where $Z$ is applied stocastically with a probability of

\begin{equation}
	p_{dephase}= \sin(\theta/2)^2.
	\label{dephasing_rate_eqn}
\end{equation}

The dephasing rate needed to account for the logical QEC cycle error rate was 0.26 Hz for the coherent simulation and 0.43 Hz for the incoherent simulation, indicating that coherent build up in the distance three color code may affect its performance and should be studied in future work.
However, we expect such single-qubit coherent noise to be less of an issue as the distance of the code is increased as suggested by \cite{iverson2019coherence}. Note that all simulation numbers reported were generated using the coherent dephasing model; however, given identical error parameters except for the dephasing rates mentioned above, the results of the two simulations were nearly identical.

\begin{table}[ht]
\begin{center}
\resizebox{\columnwidth}{!}{
\begin{tabular}{|c||c|c|c|c|}
    \hline
    \textbf{Basis} & \multicolumn{2}{c}{\textbf{SPAM}} \vline &\multicolumn{2}{c}{\textbf{QEC cycle}} \vline\\
    \hhline{|=||=|=|=|=|}
     &Sim&Exp&Sim&Exp\\
     \hline
     $\overline{X}: \; p_{yz}$ & $ 4.9(1)\times10^{-4} $ & $1.8(4)\times10^{-3}$ &$2.0(1)\times10^{-2}$&$1.89(6)\times10^{-2}$ \\
    \hline
     $\overline{Y}: \; p_{xz}$ & $4.9(1)\times10^{-4}$  &$  1.7(3)\times10^{-3}$ &$  2.74(4)\times10^{-2}$&$ 2.4(1)\times10^{-2}$\\
     \hline
      $\overline{Z}: \; p_{xy}$ & $ 4.7(1)\times10^{-4}$ &$ 1.3(3)\times10^{-3}$ &$  7.2(1)\times10^{-3}$&$1.09(3)\times10^{-2}$\\
       \hline
\end{tabular}}
\end{center}
\caption{Experimental and simulated logical SPAM error obtained from the data in Fig.~\ref{log_err_basis}. The average SPAM error rate is $1.6(4)\times10^{-3}$ for experiment and $4.9(1)\times10^{-4}$  for simulation, and the average QEC cycle error rate is $1.80(6)\times10^{-2}$ for experiment and $1.85(7)\times10^{-2}$ for simulation.}
\label{SPAM_fid_sim}
\end{table}

\begin{table}[ht]
\begin{center}
\resizebox{\columnwidth}{!}{
\begin{tabular}{|c||c|c|c|c|}
    \hline
    \textbf{Error} & \multicolumn{2}{c}{\textbf{SPAM}} \vline &\multicolumn{2}{c}{\textbf{QEC cycle}} \vline\\
    \hhline{|=||=|=|=|=|}
     &Sim&Exp&Sim&Exp\\
     \hline
     $p_x$ &  $2.3(1)\times10^{-4}$ &$  6(3)\times10^{-4}$ &$ 7.0(6)\times10^{-3}$ &$8.1(6)\times10^{-3}$\\
    \hline
     $p_y$ & $2.4(1)\times10^{-4}$ &$  6(3)\times10^{-4}$ & $2(6)\times10^{-4}$&$2.8(6)\times10^{-3}$\\
     \hline
      $p_z$ &  $2.5(1)\times10^{-4}$ &$  1.1(3)\times10^{-3}$ &  $2.76(6)\times10^{-2}$&$1.60(6)\times10^{-2}$\\
      \hline
       $p_L$ &  $7.3(4)\times10^{-4} $&$ 2.4(3)\times10^{-3}$ & $2.76(6)\times10^{-2}$&$2.70(6)\times10^{-2}$ \\
       \hline
\end{tabular}}
\end{center}
\caption{Experimental and simulated logical SPAM and QEC cycle error channel rates for the channel in Eq.~\ref{err_chan_eq} given basis error rates from Table~\ref{SPAM_fid_sim}.}
\label{err_chan}
\end{table}

\subsection{Logical error channel and error budget}

To understand how different physical error sources contribute to the logical error rate of a QEC cycle, we construct an error budget for unitary gates (the single and two qubit gates), measurement and initialization (SPAM as well as mid-circuit measurement and reset, MCMR), and dephasing noise. Note, these three error sources contain multiple noise mechanisms. For example, the unitary gate errors contain not just depolarizing noise from the single and two qubit unitary gates but also spontaneous emission, which includes leakage, as well (see the supplemental material for more details). Given the error model of the simulation, the error budget captures how the noise from the three error sources impacts the logical error rate of a QEC cycle. This is particularly useful since it can be unclear how errors such as leakage and dephasing translate from the physical to the logical system.

\begin{figure}
\includegraphics[width=\columnwidth]{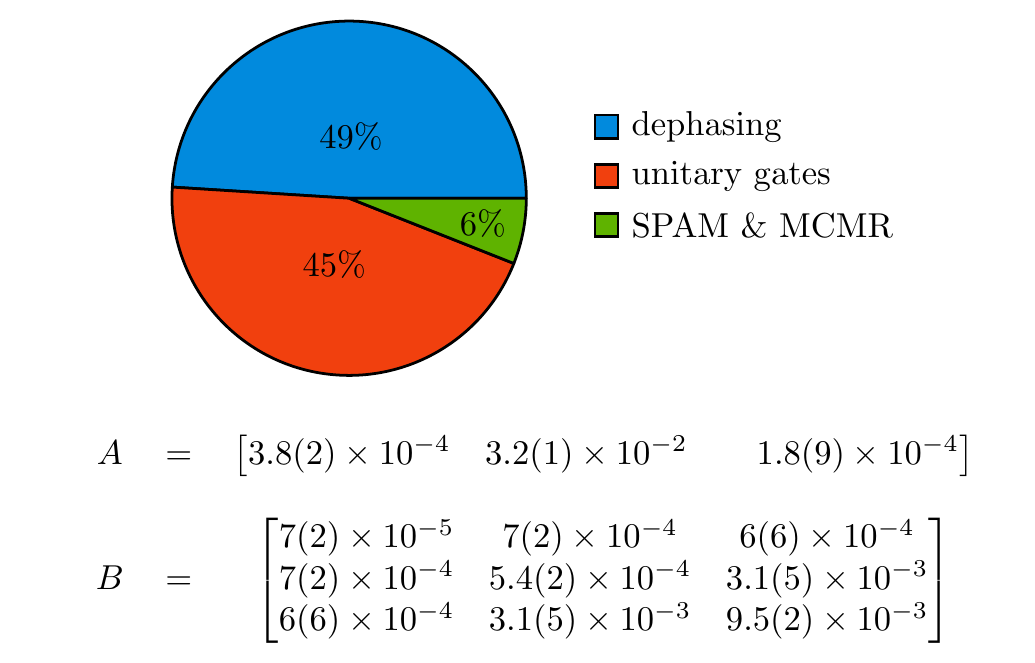}
\caption{The logical qubit error budget. The pie chart shows the percent contribution to the logical QEC cycle error rate from the physical noise due to SPAM and mid-circuit measurement and reset (MCMR), unitary gates, and qubit dephasing. $A$ and $B$ are the vector and matrix in Eq.~\ref{error_budget_eqn} and the elements are ordered as $\{SPAM/MCMR, gates,dephasing\}$. The individual contributions in the pie chart indexed by $i$ are calculated as $(A_i+\sum_jB_{ij})/(\sum_iA_i+\sum_{ij}B_{ij})$.
}
\label{error_pie}
\end{figure}

The error budget was calculated via simulation assuming a second-order model given by

\begin{equation}
p_L(s)=s\sum_ia_ip_i+s^2\sum_{i,j}b_{ij}p_ip_j
= s\sum_i A_i+s^2\sum_{i,j}B_{i,j},
\label{error_budget_eqn}
\end{equation}
where $s$ is a scaling factor used for fitting, $p_i$ are the relative probabilities of SPAM, two-qubit gate, and dephasing errors at the physical level, and $A_i = a_ip_i$ and $B_{ij} = b_{ij}p_ip_j$ are the coefficients we solve for to determine the relative impact of the physical layer errors on the logical qubit. We first set two of the three errors to zero and then, for each error, scale $s$ in simulations to determine $A_i$ and $B_{ii}$ using three different quadratic curve fits. The off-diagonal elements of $B$ are then solved for by only setting one error to zero and setting $s=1$.
The estimated individual and correlated contributions to the logical error are shown in Fig.~\ref{error_pie}.

The error budget breakdown indicates that the noise due to unitary gates and dephasing account for the majority of the logical error, accounting for approximately $49\%$ and $45\%$, respectively, while measurement and initialization account for the remaining $6\%$. However, studying only the individual contributions, obtained by the calculation in Fig.~\ref{error_pie}, we miss some important information contained in $A$ and $B$. As can be seen in Fig.~\ref{error_pie}, the contribution to the logical error that is linear in the physical errors, $A$, is dominated by the gate errors. Ideally, the logical error probability should be quadratic in the gate error, and the large linear dependence is likely due to the leakage error associated with the gate operations. This point will be further illustrated below as we study the pseudo-threshold of the code.

\subsection{Pseudo-threshold estimates}

While the error budget is useful in studying the current level of impact different physical error sources have on the QEC protocol, we now investigate a simulation tool better suited to predicting the protocol performance assuming improved physical error rates.

For QEC to be helpful in computations, logical level error rates must be below the physical level error rates.
This crossover point is known as the `pseudo-threshold'.
In general, pseudo-threshold estimates are difficult because they require a detailed understanding of the system's underlying physics.
However, in cases where the dominant physical noise mechanism is known, one method is to scale that noise mechanism in simulations and solve for the pseudothreshold, defining it as the point where the logical error rate is equal to the dominant physical noise source.
The largest error rate in the simulation model is the two-qubit depolarizing error rate of $p2 = 3.1 \times 10^{-3}$ (see supplementary material for details). Note that the two-qubit gate not only contributes depolarizing noise, but also leakage noise via spontaneous emission at a probability of $5.5 \times 10^{-4}$; however, to simplify discussions we consider only the two-qubit depolarizing rate. Also, while dephasing has a large impact on the logical error budget, the physical level dephasing error rate per physical qubit operation is an order of magnitude lower than $p2$. That is, with a dephasing rate of $0.26$ Hz and each duration between gates, the stochastic error probability averaged over the entire circuit is approximately $2.2 \times 10^{-4}$ for each qubit. Thus, in the simulations, on the physical per operation level, the dephasing error channel has a lower error rate than the two-qubit depolarizing rate.

\begin{figure}
	\includegraphics[width=\columnwidth]{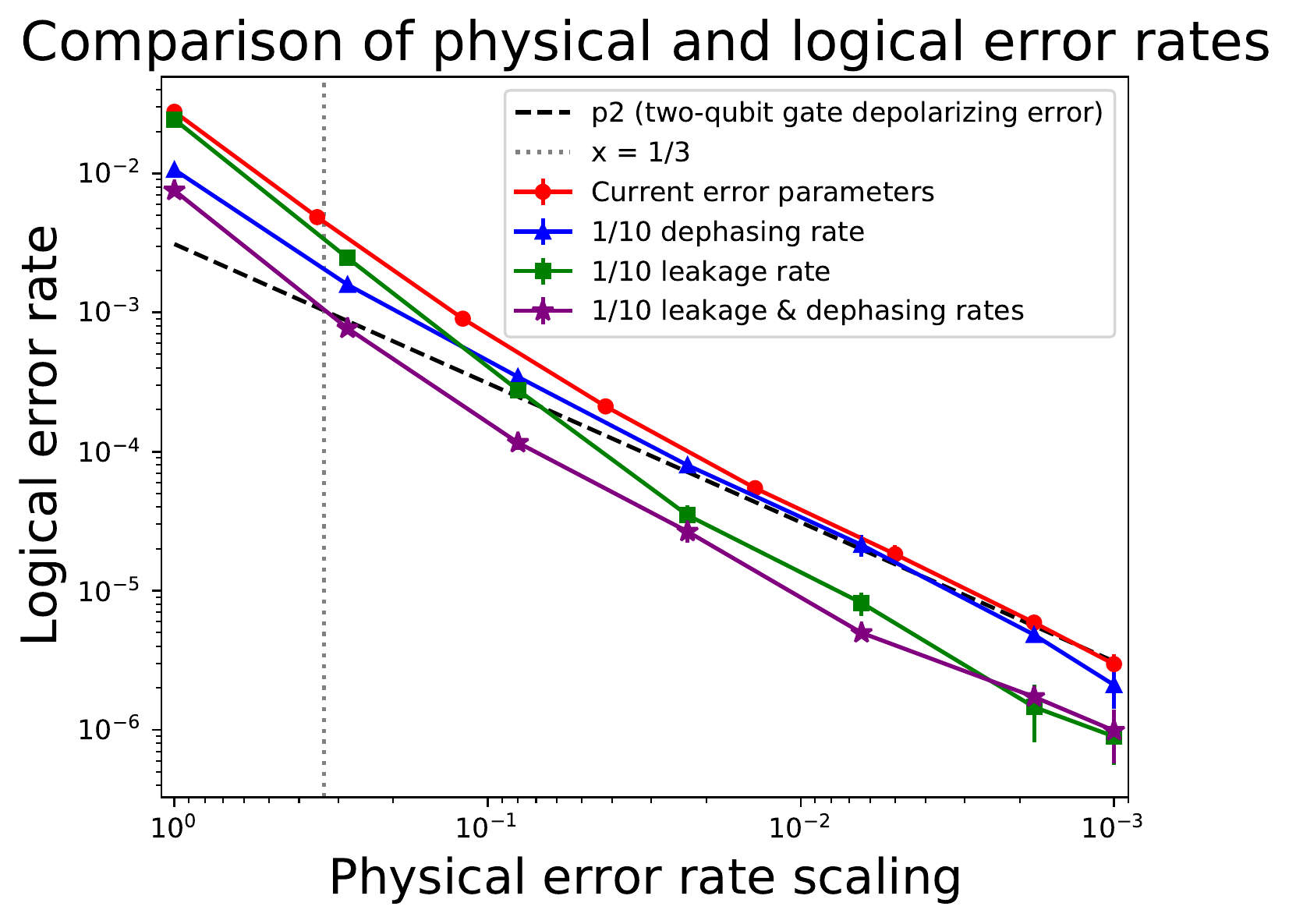}
	\caption{Comparison of the logical error rate to different error models as the physical error rates are scaled. The dashed line is equal to the physical layer two-qubit depolarizing error rate, $p2$, times the scaling factor defining the x-axis, which we use to define our pseudo-threshold line. The four solid lines represent the logical qubit error rate for different simulated error model scenarios and are plotted as a function of an overall scaling of the error model parameters. The red line (circles) uses our current error model parameters estimate detailed in Table~\ref{Sim_Parameters}, the blue line (triangles) was generated assuming our current error parameters except with the dephasing rate being reduced by 10x, the green line (squares) was generated assuming our current error parameters except with the leakage rate being reduced by 10x, and the purple line (starts) was generated by starting with our current error parameters and suppressing both the dephasing and the leakage rates by 10x. The vertical dotted line ($x = 1/3$) is approximately where the purple line crosses the pseudo-threshold.  We find that reducing the leakage will be key to achieving a QEC cycle logical error rate below pseudo-threshold (when the logical error rate is less than the leading physical error rate). Note, some error bars are smaller than the markers.}
\label{pseudo_thres}
\end{figure}

While our simulations lack the level of detail needed for precise estimates, we use a crude model to estimate system improvements needed to pass the conjectured pseudo-threshold.
To this end, we first simulated the logical error rate of a single QEC cycle and scale all physical error rates used (see Table \ref{Sim_Parameters}) by a common scaling factor.
This is a coarse view of how much our dominant bare physical error (the two-qubit gate) needs to improve to reach the simulated pseudo-threshold.
As seen in Fig. \ref{pseudo_thres}, simply improving all our errors by a common factor is not enough to reduce our logical error rate below the two-qubit depolarizing error rate $p2$.
To understand this, we also simulated three other models with different relative error budget breakdowns, first reducing the dephasing rate to $1/10$th the original value, then reducing the leakage rate to $1/10$th the original value  and then both reducing the dephasing and leakage rate to $1/10$th their original value.
By scaling these additional error models by an overall scaling factor, we find that suppressing the dephasing rate is likely not sufficient to reach the pseudo-threshold, but that substantially reducing the relative leakage rate \textit{and} the dephasing rate should put the pseudo-threshold within reach if the two-qubit gate error can be reduced by approximately a factor of 3.

The relative importance of taming leakage errors is not unexpected.
Most QEC protocols were designed to correct errors within the qubit manifold, an assumption that leakage errors violate.
When leakage errors occur (\textit{e.g.}, during spontaneous laser scattering events), leaked qubits can spread errors as they the interact with other qubits and the circuits for the color code are in fact only FT to qubit errors and not leakage errors.
While leakage is not a leading source of error in our system, if left untreated it will saturate the system, eventually corrupting all the qubits and the logical information.
As shown in Ref.~\cite{hayes2020eliminating}, leakage errors can be reduced by roughly three orders of magnitude through physical mitigation techniques. Alternatively, circuit level techniques can be incorporated in the QEC procotol~\cite{suchara2015leakage, brown2020critical}.

Additionally, we believe there are routes to improving dephasing rates, including the use of dynamical decoupling, additional shielding and spatial phase tracking in the control software to account for spatial inhomogeneities in magnetic field.

\section{Conclusion}
In this work, we demonstrated the primitives needed for quantum error correction restricted to a single logical qubit, including high fidelity state-preparation and readout of logical basis states and a magic state, logical single-qubit gates, and repeatable error correction cycles.
By establishing the necessary hardware capabilities along with detailed simulations of the processes, we can now begin developing a QCCD system architecture that is optimized for computations at the logical level.

The experimental data and simulated results highlight the need for substantial error rate improvements to get well below the pseudo-threshold.
The largest contributor to the physical level error budget is the two-qubit gate and, as noted in Ref. \cite{Pino2020}, we believe the dominant error mechanisms can be suppressed with upgraded electrode voltage sources and laser systems.
Using the simulator developed here, we identify leakage and dephasing as crucial sources of noise. Interestingly, while leakage is not currently a dominant physical error, scaling the physical error rates in simulation indicates that leakage will dominate the logical error rate.
While leakage errors are particularly detrimental to code performance, they can be converted to Pauli errors with the addition of a leakage repumping routine~\cite{hayes2020eliminating} and should not present a fundamental roadblock.
Perhaps surprisingly, even though dephasing errors are relatively small compared to two-qubit gate errors at the physical level, our logical qubit error budget indicates dephasing errors have a large impact on the code performance.
Fortunately, trapped-ion hyperfine qubits have been shown to exhibit much longer coherence times~\cite{Wang21} compared to our current dephasing rates, and we believe this particular error can be significantly improved by a combination of improved shielding, dynamical decoupling, and accurate mapping of magnetic field inhomogeneities and spatial phase tracking in the control software. Additionally, dephasing errors can be suppressed further through speed improvements, which are currently bottlenecked by laser cooling and transport operations, both of which can be substantially improved through advanced techniques~\cite{Jordan2019,Mourik20}, and we note that smaller trap geometries~\cite{Ivory20} or 2-dimensional topologies~\cite{Wright_2013} could also improve transport times. At the logical level, some interested noise mitigation techniques include Pauli or Clifford twirling, randomized compiling, circuit-level gadgets to suppress coherent errros~\cite{debroy2018stabilizer, zhang2021hidden, Parrado_Rodr_guez_2021} and leakage errors~\cite{suchara2015leakage, brown2020critical}, subsystem codes, codes designed for biased noise~\cite{Tuckett18}, and larger distance codes~\cite{iverson2019coherence}.

These experiments and emulation tools pave the way for co-designing QCCD systems and QEC software to implement large-scale fault-tolerant quantum computers. The next milestones on the road to a  fully error corrected quantum computer include logical operations between multiple qubits \cite{erhard2020entangling, landahl2014quantum}, and operation below the pseudo-threshold.

\section{Acknowledgements}
We would like to thank Charlie Baldwin and Michael Foss-Feig for helpful comments on the manuscript and Daniel Lidar for useful discussion. Most importantly, we thank the entire group at HQS for their superb work and many contributions. Particularly, we thank Joe Chambers and Tom Skripka for the real-time engine implementation capable of making fast control decisions based on mid-circuit measurement results, we thank Raanan Tobey and Matt Bohn for specialized laser system work enabling high-fidelity gate operations, and we thank Bryan Spann for his work to construct parallelized laser beam delivery systems.

\bibliography{inversion}
\appendix

\section{Methods\label{appen:method}}

The physical qubits are encoded in the $^{171}Yb+$ $S_{1/2}$ hyperfine clock states $\ket{F=0,m_f=0}$ and $\ket{F=1,m_f=0}$ with $F$ and $m_f$ respectively being the total angular momentum and the z-projection quantum numbers.
Gating operations are implemented with stimulated Raman transitions ~\cite{Pino2020,Baldwin2019} with single-qubit gates being performed with 2-ion $\{\mathrm{Yb,Ba}\}$ crystals using two co-propagating circularly polarized beams at 368.0 nm, and two-qubit operations being performed with 4-ion $\{\mathrm{Yb,Ba,Ba,Yb}\}$ crystals using two beams with a wave-vector difference $\Delta k$ coupling to the axial motion.
The single $^{171}$Yb$^+$ axial center-of-mass mode frequency is 1.0 MHz and we use the first higher-order mode at 1.74 MHz for entangling operations.
Before any two-qubit gate operations, motional excitations are cooled out of the system using a combination of Doppler and resolved sideband cooling on the $^{138}$Ba$^+$ ions.

We characterize physical qubit operations using error rates extracted from parallel randomized benchmarking ~\cite{Magesan2011}.
Physical qubits are initialized and measured using standard optical pumping and state-dependent fluorescence techniques~\cite{Olmschenk07} with an average SPAM error of $2.4(9) \times 10^{-3}$, where the uncertainty indicates the typical variations across the zones and over time. During qubit measurement and reset, laser scatter and ion fluorescence absorbed by idle qubits cause errors ranging from $<1.0\times10^{-4}$ to $3\times10^{-3}$ depending on the measurement zone and the other qubits' physical locations. Single-qubit and two-qubit gate errors are respectively measured to be $7(1)\times10^{-5}$ and $3.0(1)\times10^{-3}$. The memory error in a depth-one circuit is benchmarked at $< 5(3)\times10^{-4}$. Our estimates for errors in transport operations, such as physical swap, is negligible compared to other errors.

We compile OpenQASM to a hardware-specific pulse language that combines electrode and laser control. The sequences are executed with AWG’s driving electrode voltages and DDS components controlling laser pulses, both sequenced by a distributed network of controllers which combine programmable logic components with ARM microprocessors.  The microprocessor allows for sequencing the error corrections as optional laser pulses embedded in a fixed ion transport schedule. To execute conditional gates, qubit measurements are analyzed and broadcast by a single microprocessor. Typically, the conditional information is distributed in parallel with unconditional circuit operations, and in worst case conditions, adds $<$250 $\mu$s of distribution latency.

\begin{figure*}[ht!]
	\includegraphics[trim=0 0 0 0, clip, scale=0.75]{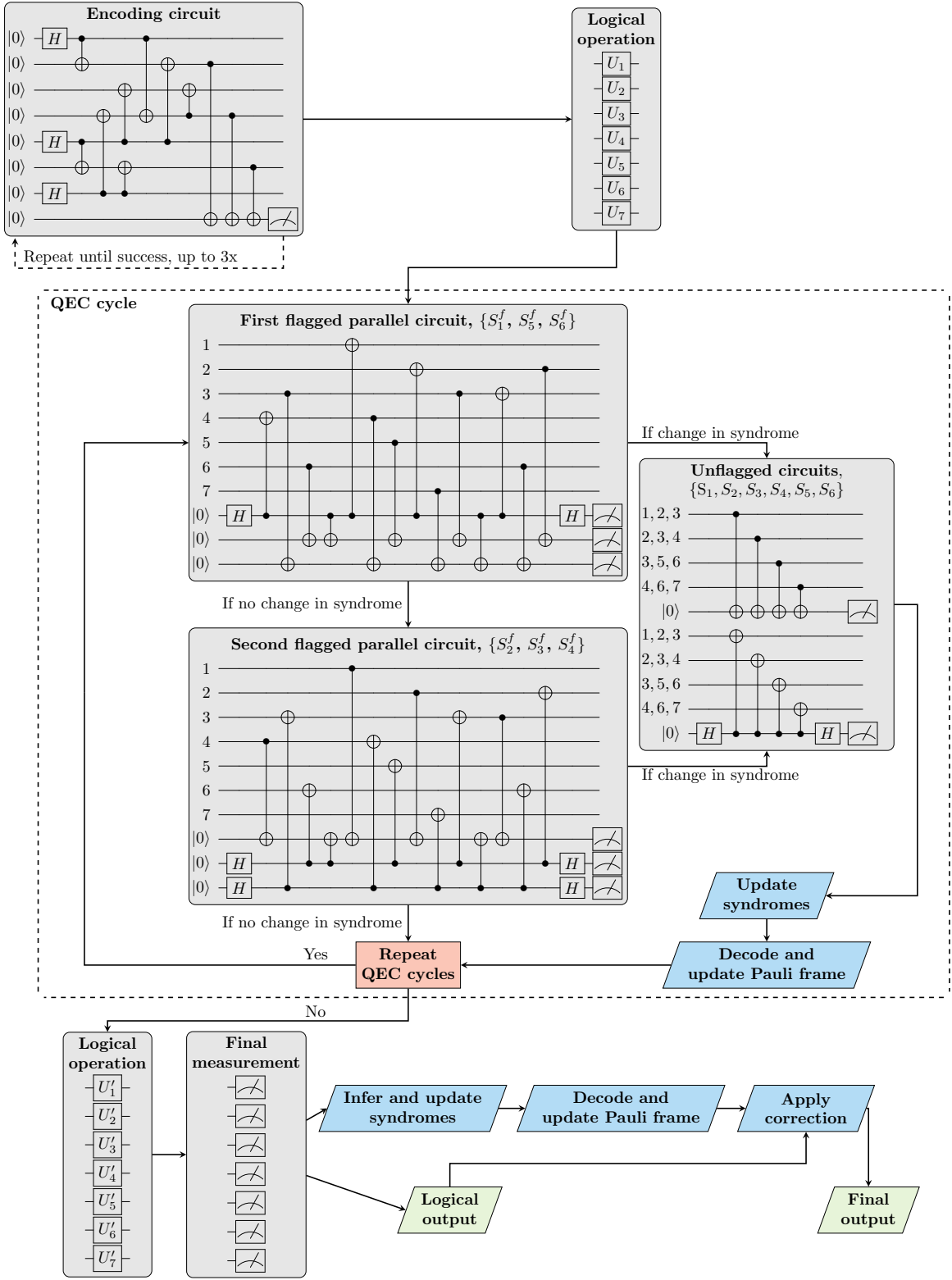}
	\caption{A detailed schematic of the QEC demonstration. The encoding circuit first prepares $\ket{0}_L$ and uses an ancilla to verify this preparation. If the ancilla measured is found in $\ket{0}$, the circuit succeeded and moves the next step. If the ancilla is found to be in $\ket{1}$, then all qubits are reinitialized and the circuit ran again until successful, up to three times. After a maximum of three initialization attempts, we proceed by applying single qubit unitaries to make a logical rotation to prepare the desired logical state. Adaptive QEC cycles are then performed. The first set of three syndromes is measured using the flag circuit $\{S^f_1, S^f_5, S^f_6\}$. If these syndromes do not indicate an error, then the second set of three sydromes is measured $\{S^f_1,  S^f_3,  S^f_4\}$ and if no errors are indicated, the syndrome extraction protocol is complete. If either of the flagged circuits do indicate an error, the protocol moves to measuring all six stabilizers using the unflagged circuit $\{S_1,S_2,S_3, ... \}$, thus completing the syndrome extraction. The syndrome information is then fed to the decoder to infer a correction and the Pauli frame is updated.  Finally, the state is rotated to the appropriate basis for measurement using single qubit unitaries. Upon measurement, syndrome information is inferred and sent to the decoder for a final correction.}
	\label{FlowChart}
\end{figure*}
\begin{figure*}
	\includegraphics[trim=0 0 10 0, clip, width=\textwidth]{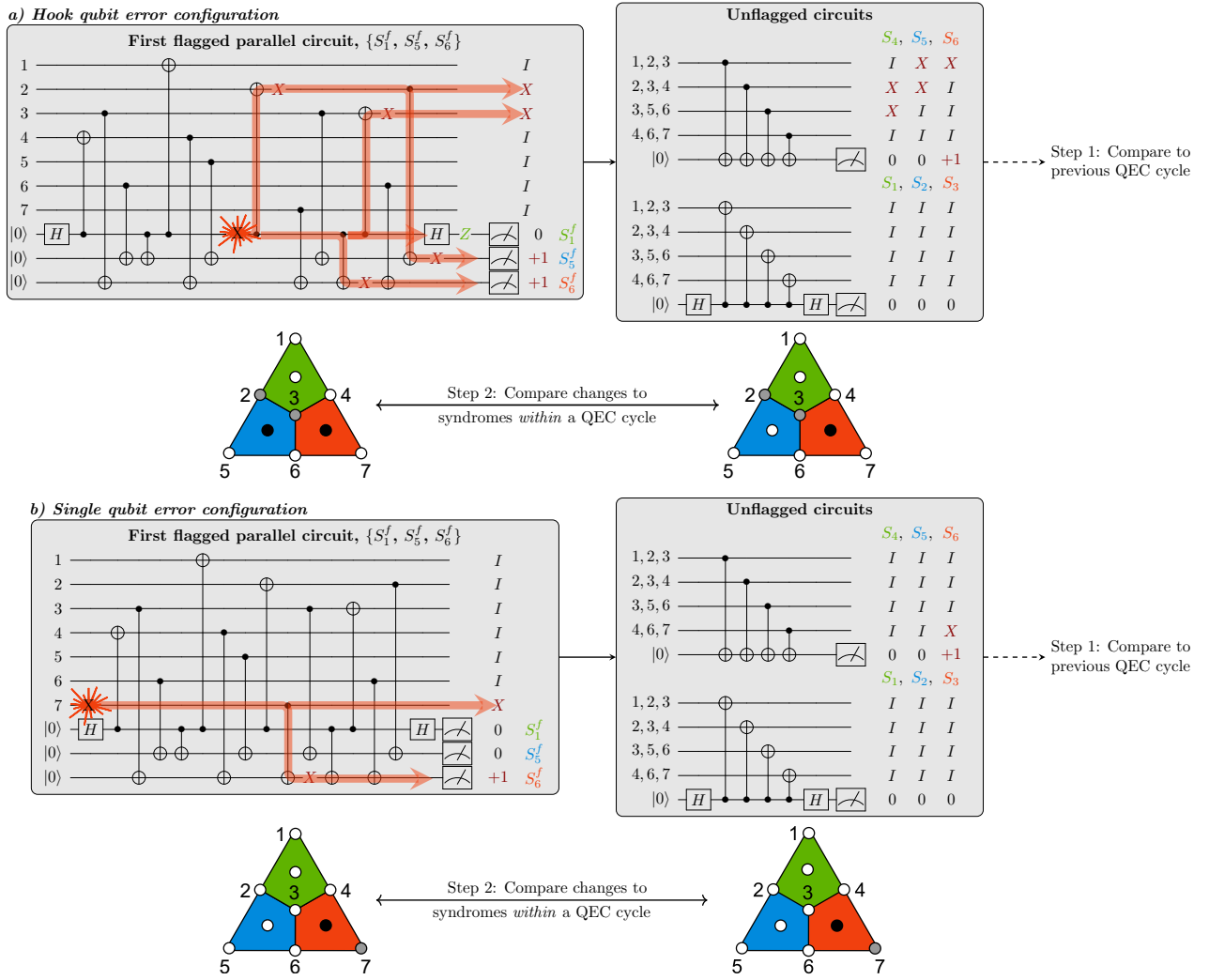}
	\caption{An outline of the decoder procedure and two different error configurations. \textit{a)} A single qubit hook error occurs on an ancilla and spreads to two data qubits. This triggers the additional round of syndrome extraction using the unflagged circuits. We denote the errors on the data qubits (gray circles) and the changed syndromes measurements (black circles) pictorially on the color code figure. \textit{b)} A single qubit error occurs on a data qubit and causes a change in the syndrome measurements, again, triggering an additional round of syndrome extraction. In both \textit{a)} and \textit{b)} the final round of syndrome measurements are identical, meaning the first step of the decoder cannot distinguish between these two error configurations. The second step of the decoder compares the syndromes measured within the QEC cycle to distinguish between cases such as this. We see that the two sets of measured syndromes are different for the hook error, but the same for the single qubit data error. }
	\label{hook_decoder}
\end{figure*}
\begin{figure*}
	\includegraphics[angle=270, trim=0 0 420 0, clip, scale=0.75,width=\textwidth]{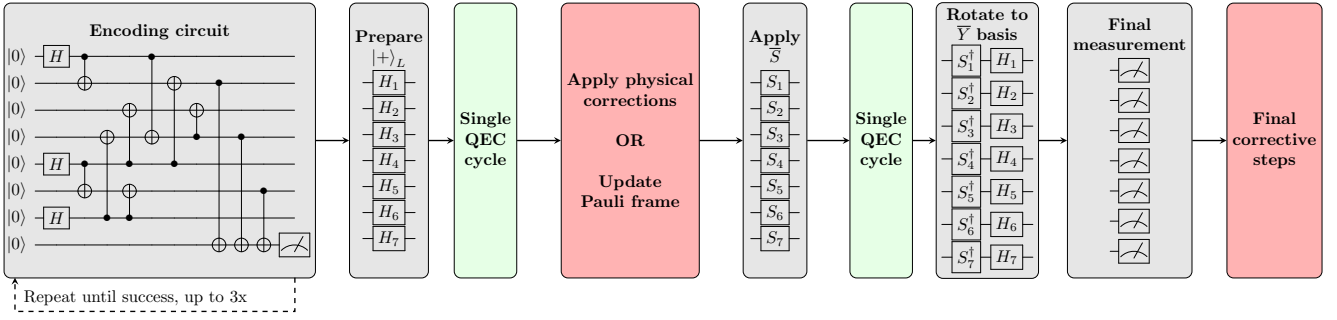}
	\caption{A schematic of the $\overline{S}$ gate experiments. In this set of experiments, we compare physically applied corrections to implementing the corrections in software using the Pauli frame. The main elements are the same as those used in Fig. \ref{FlowChart}. First, using the FT encoding circuit, we prepare $\ket{0}_L$ and rotate to $\ket{+}_L$ with an $\overline{H}$ gate. We then perform one QEC cycle to generate corrections. Next, we either physically apply corrections to the qubits, or update the Pauli frame in software, followed by physically applying the $\overline{S}$ gate. We then apply one last QEC cycle to generate further corrections, rotate to the $\overline{Y}$ basis, and measure the data qubits. The final corrective steps infer the syndromes, decode, and apply corrections to the output data as outlined in the main text and Fig. \ref{FlowChart}.}
	\label{S_gate}
\end{figure*}
\begin{figure}
	\includegraphics[trim=0 650 0 0, clip, scale=0.75]{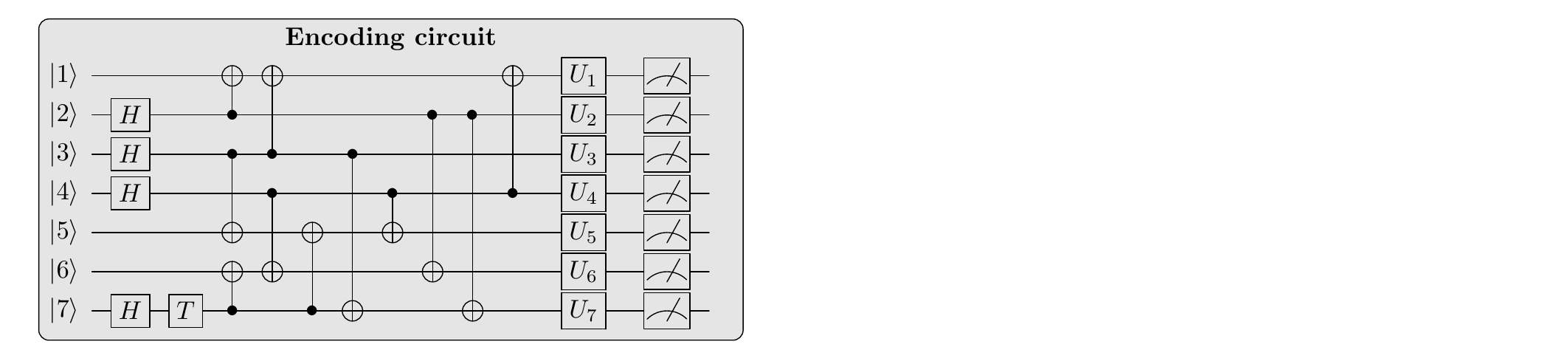}
	\caption{A non-fault-tolerant encoding circuit used to prepare the magic state $\overline{T}\ket{+}_L$. }
	\label{TState}
\end{figure}

\subsection{Simulations}
The simulator used in this work is a modified version of the software PECOS~\cite{pecos} that receives instructions directly from compiler, which is composed of the native quantum gate set, classical operations, and transport operations. That is, the instructions that were used to encode the QEC experiments in this paper and were executed by the simulator were same instructions that QCCD device was instructed to perform (before being translated to hardware-specific pulses). The simulations models errors both coherently using a state-vector backend \cite{steiger2018projectq} as well as incoherently using a stabilizer simulation \cite{crathesis}. All simulation results presented in this paper are those of the coherent simulation; however, both the coherent and incoherent simulations produce nearly identical results given a dephasing rate of $0.25$ Hz and $0.43$ Hz, respectively.

The error model includes simple depolarizing gate noise, leakage errors, and dephasing noise during transport and cooling operations.
Errors on physical qubits are modeled as stochastic processes (excluding dephasing). Most errors are applied with probabilities determined by independent experiments as summarized in Table \ref{Sim_Parameters}, with the exception of dephasing errors. The coherent simulation modeled dephasing as $RZ(\theta)$ following ideal gates, where $\theta$ is the dephasing rate, $0.26$ Hz times the time between gates due to qubit idling or transport operations. For the incoherent simulations, dephasing errors are applied as stochastic $Z$ errors between gate operations with a probability of $p=sin(\theta/2)^2$. That is, probability obtained from Pauli twirling the coherent channel. Note here, the same $theta$ is used as the coherent model. Both dephasing rates ($0.26$ Hz and $0.43$ Hz) were empirically found, as discussed in the main text.

Physical qubit initialization to $\ket{0}$ occurs at the beginning of each circuit and after each measurement. The fidelity of this procedure is limited by off-resonant coupling which is numerically simulated. The residual population remains in the $F=1$ manifold, $2/3$'s of which is distributed in the leakage states $\ket{F=1,m_f=\pm1}$. The single-qubit and two-qubit gate errors are modeled as being dominated by a depolarization process whose amplitude is measured via randomized benchmarking experiments. The spontaneous emission that occurs during stimulated Raman transitions is estimated through atomic physics calculations closely following Ref.~\cite{Ozeri07}. While the spontaneous emission is estimated to be small compared to the total error, the leakage induced by the process is more detrimental than errors that keep the ion in the qubit subspace, and we therefore explicitly include this error in addition to the depolarizing error. Note that the spontaneous emission process is modeled as causing leakage with probability $1/2$, and causing $X$ and $Y$ errors each with probability $1/4$. Single-qubit rotations about the z-axis are done entirely in software, and therefore contribute negligible errors. We model measurement errors as bit flips and note that leaked qubits return measurement results that are indistinguishable from $\ket{1}$ due to the measurement process described in Ref.~\cite{Olmschenk07}.
\begin{table}[]
\begin{tabular}{|l|l|l|}
\hline
\textbf{Operation}                       & \textbf{Channel}                   & \textbf{Probability}             \\ \hhline{|=|=|=|}
\multirow{2}{*}{Initialization}        & Bit flip                  & $1.66\times10^{-6}$                   \\ \cline{2-3}
                                       & Leakage                   & $3.33\times10^{-5}$                   \\ \hline
\multirow{2}{*}{X/Y Single-qubit gate} & Depolarizing              &  $7\times10^{-5}$                       \\ \cline{2-3}
                                       & Spontaneous Emission      &  $1.25\times10^{-5}$                       \\ \hline
\multirow{2}{*}{Two-qubit gate}        & Depolarizing              &  $3.1\times10^{-3}$                       \\ \cline{2-3}
                                       & Spontaneous Emission      & $ 2.75\times10^{-4}$                       \\ \hline
\multirow{2}{*}{Measurement}           &{Bit flip} & {$2.4\times10^{-3}$} \\ \cline{2-3}
                                       &     Leakage                      &              $5\times10^{-3}  $         \\ \hline
\multirow{2}{*}{Crosstalk}        &Initialization                  & $2.3\times10^{-5}$                   \\ \cline{2-3}
                                       &Measurement                   & $2.3\times10^{-4}$      \\ \hline \hline
\textbf{Operation}                       & \textbf{Channel}                   & \textbf{Rate (Hz)}             \\ \hhline{|=|=|=|}
\multirow{2}{*}{Dephasing}        &Coherent                  & $ 0.26$                   \\ \cline{2-3}
                                       &Incoherent                   & $0.43$      \\ \hline
\end{tabular}
\caption{Simulation error parameters. With the exception of the dephasing rate, the parameters used come from either experimental measurement or well defined microsopic noise analysis. Values based on microscopic noise analysis include further break-down of the errors such as initialization error consisting of 1/3 bit-flip errors and 2/3 leakage, and spontaneous emission consisting of 1/2 leakage error, 1/4 $X$ error, and 1/4 $Z$ error. The dephasing rates used in the simulations were determined separately for the incoherent and coherent simulations by adjusting the dephasing rate until the logical error rate for a QEC cycle matched the value found from the experiment.}
\label{Sim_Parameters}
\end{table}%

\begin{figure*}[ht!]
\begin{minipage}[t]{\textwidth}
\begin{lstlisting}[label=code:qec-cycle_exp, caption={The main program for the repeated QEC cycle and the magic-state experiment.},style=pseudo,]
def qec_cycles_exp(init_state: str, meas_basis: str, num_cycles: int, shots: int) -> Tuple[int, int]:
	"""
	Pseudocode representing the repeated QEC cycle and the magic-state experiments.

	Args:
		init_state: The logical state to initialize.
		meas_basis: The logical measurement basis to measure in the end.
		num_cycles: The number of QEC cycles.
		shots: The number of experiments to attempt.

	Returns:
		A tuple of the number of shots and the number of shots with measurement outcomes consistent with state preparation.
	"""
	runs = 0     # Number of total shots
	success = 0  # Number of shots that had expected outcomes

	for i in range(shots):

		# Initialize the logical state
		if init_state == '|T>':
			# Encode %\textcolor{MediumGray}{$T\ket{+}$}% using Steane code encoding circuit from Ref.%\textcolor{MediumGray}{~\cite{preskill1998reliable}}%
			state = prep_tstate()
		else:
			# Use FT circuit from Ref.%\textcolor{MediumGray}{~\cite{goto2016minimizing}}% to prepare logical %\textcolor{MediumGray}{$\ket{0}$}% (see Fig.%\textcolor{MediumGray}{~\ref{FlowChart}}%)
			state = ftprep_zero()  # prepare state with up to 3 attempt to verify, continue regardless of final verification

			# Rotate that state from logical |0> to another logical state as apporpriate
			state = logical_rotate_init_basis(state, init_state)

		# Run the QEC cycles
		last_syndromes_x = [0, 0, 0]
		last_syndromes_z = [0, 0, 0]
		pf = [0, 0]  # The Pauli frame: [Apply logical X?, Apply logical Z?]

		for j in range(num_cycles):
			state, last_syndromes_x, last_syndromes_z, pf = qec_cycle(state, last_syndromes_x, last_syndromes_z, pf)

		# Rotate the logical measurement basis
		state = rotate_meas_basis(state, meas_basis)

		# Do a destructive logical measurement
		meas_output = logical_meas(state, meas_basis, last_syndromes_x, last_syndromes_z, pf)

		runs += 1

		# Determine if measurement result was as expected. If so, upade success count.
		expected_outcome = expected_result(meas_output, init_state, meas_basis)
		success += expected_outcome

	return runs, success
\end{lstlisting}
\end{minipage}%
\end{figure*}%

\begin{figure*}
\begin{minipage}[t]{\textwidth}
\begin{lstlisting}[label={code:sgate}, caption={The main program for the active vs software correction experiment with the logical S.},style=pseudo,]
def active_corr_exp(active_correction: bool, shots: int) -> Tuple[int, int]:
	"""
	Pseudocode representing the active vs software experiments.

	Args:
		active_correction: Whether to physically apply the logical corrections before the logical S gate or not.
		shots: The number of experiments to attempt.

	Returns:
		A tuple of the number of shots and the number of shots with measurement outcomes consistent with state preparation.
	"""
	runs = 0     # Number of total shots
	success = 0  # Number of shots that had expected outcomes

	for i in range(shots):

		# Initialize the |+> ogical state
		# Use FT circuit from Ref.%\textcolor{MediumGray}{~\cite{goto2016minimizing}}% to prepare logical %\textcolor{MediumGray}{$\ket{0}$}% (see Fig.%\textcolor{MediumGray}{~\ref{FlowChart}}%)
		state = ftprep_zero() # prepare state with up to 3 attempt to verify, continue regardless of final verification

		# Rotate that state from logical |0> to logical |+>
		state = logical_rotate_init_basis(state, init_state='|+>')

		# Run the QEC cycles
		last_syndromes_x = [0, 0, 0]
		last_syndromes_z = [0, 0, 0]
		pf = [0, 0]  # The Pauli frame: [Apply logical X?, Apply logical Z?]

		# Do a single QEC cycle
		state, last_syndromes_x, last_syndromes_z, pf = qec_cycle(state, last_syndromes_x, last_syndromes_z, pf)

		if active_correction:  # Physically apply any logical X or Z corrections
			# We apply both X and Z; however, you don't have to apply Z since S and Z commute
			state = apply_pauli_frame(state, pf)
			pf = [0, 0]  # Reset Pauli frame
		else:
			# Apply the S gate rotation to the Pauli frame
			# Z -> Z, X -> Y
			pf = sgate2pf(pf)  # equivalent to XORing: pf[1] = pf[1] ^ pf[0]

		# Physically apply the logical S gate
		state = logical_sgate(state)

		# Rotate the logical measurement basis to Y since S: |+> -> |+i>
		state = rotate_meas_basis(state, meas_basis='Y')
		# Do a destructive logical measurement
		meas_output = logical_meas(state, meas_basis, last_syndromes_x, last_syndromes_z, pf)

		runs += 1

		# Determine if measurement result was as expected. If so, upade success count.
		expected_outcome = expected_result(meas_output, init_state, meas_basis)
		success += expected_outcome

	return runs, success
\end{lstlisting}
\end{minipage}%
\end{figure*}%

\begin{figure*}
\begin{minipage}[t]{\textwidth}
\begin{lstlisting}[label={code:rotate-state}, caption={Rotate logical $\ket{0}$ to the appropriate state.},style=pseudo,]
def logical_rotate_init_basis(state: QuantumState, init_state: str) -> QuantumState:
	"""
	Pseudocode representing rotating to appropriate logical basis.

	Args:
		state: The logical state to rotate.
		init_state: The logical state to rotate to.

	Returns:
		A rotated logical state.
	"""

	if init_state == '|0>':
		# |0> -> |0>
		pass  # do nothing

	elif init_state == '|1>':
		# X |0> -> |1>
		state = logical_x(state)  # X on qubits 5, 6, and 7  (|0> -> |1>)

	elif init_state == '|+>':
		# H |0> -> |+>
		state = logical_h(state)  # %\textcolor{MediumGray}{$H$}% on all data qubits

	elif init_state == '|->':
		# H X |0> -> |->
		state = logical_x(state)  # %\textcolor{MediumGray}{$X$}% on qubits 5, 6, and 7 (|0> -> |1>)
		state = logical_h(state)  # %\textcolor{MediumGray}{$H$}% on all data qubits (|1> -> |->)

	elif init_state == '|+i>':
		# S H |0> -> |+i>
		state = logical_h(state)  # %\textcolor{MediumGray}{$H$}% on all data qubits (|0> -> |+>)
		state = logical_s(state)  # %\textcolor{MediumGray}{$S^\dagger$}% on all data qubits (|+> -> |+i>)

	elif init_state == '|-i>':
		# S H X |0> -> |-i>
		state = logical_x(state)  # %\textcolor{MediumGray}{$X$}% on qubits 5, 6, and 7  (|0> -> |1>)
		state = logical_h(state)  # %\textcolor{MediumGray}{$H$}% on all data qubits  (|1> -> |->)
		state = logical_s(state)  # %\textcolor{MediumGray}{$S^\dagger$}% on all data qubits  (|-> -> |-i>)

	return state
\end{lstlisting}
\end{minipage}%
\end{figure*}%

\begin{figure*}
\begin{minipage}[t]{\textwidth}
\begin{lstlisting}[label=code:qec-cycle, caption={A QEC cycle},style=pseudo,]
def qec_cycle(state: QuantumState, last_syndrome_x: List[int, int, int], last_syndrome_z: List[int, int, int], pf: List[int, int]) -> Tuple[QuantumState, List[int, int, int], List[int, int, int], List[int, int]]:
	"""
	Pseudocode representing a single QEC cycle.

	Args:
		init_state: The logical state to initialize.
		meas_basis: The logical measurement basis to measure in the end.
		num_cycles: The number of QEC cycles.
		shots: The number of experiments to attempt.

	Returns:
		A logical state, the last X-type syndromes measured (not including flagged syndromes), the last Z-type syndromes measured (not including flagged syndromes), and the updated Pauli-frame, which tracks whether to apply a logical X and/or Z correction.
	"""
	flag_diff_x = [0, 0, 0]
	flag_diff_z = [0, 0, 0]

	# Measure the first set of stabilizers in parallel.
	fx0, fz1, fz2 = meas_flagging_syndromes_xzz(state)
	# XOR with previously measured syndrome to get the difference
	flag_diff_x[0] = fx0 ^ last_syndrome_x[0]
	flag_diff_z[1] = fz1 ^ last_syndrome_z[1]
	flag_diff_z[2] = fz2 ^ last_syndrome_z[2]

	# If no change detected, go on to the next flagging syndromes
	if flag_diff_x == [0, 0, 0] and flag_diff_z == [0, 0, 0]:
		# Measure the second set of stabilizers in parallel.
		fz0, fx1, fx2 = meas_flagging_syndromes_zxx(state)
		# XOR with previously measured syndrome to get the difference
		flag_diff_z[0] = fz0 ^ last_syndrome_z[0]
		flag_diff_x[1] = fx1 ^ last_syndrome_x[1]
		flag_diff_x[2] = fx2 ^ last_syndrome_x[2]

	# If any change in syndromes was detected, re-measure the set of six stabilizer generators without flagging
	if flag_diff_x != [0, 0, 0] or flag_diff_z != [0, 0, 0]:
		sx0, sx1, sx2, sz0, sz1, sz2 = meas_six_syndromes(state)
		syndromes_x = [sx0, sx1, sx2]
		syndromes_z = [sz0, sz1, sz2]
		# Get the change in sydromes
		syndrome_diff_x = bitwise_xor(syndromes_x, last_syndrome_x)
		syndrome_diff_z = bitwise_xor(syndromes_z, last_syndrome_z)

		# Determine a Pauli-frame update to track correction in software
		pf_new_x = decoder_2d(syndrome_diff_x)
		pf_new_z = decoder_2d(syndrome_diff_z)
		# Now modify the 2D decoder corrections based on flagging
		pf_flag_new_x = decoder_flag_update(syndrome_diff_x, flag_diff_x)
		pf_flag_new_z = decoder_flag_update(syndrome_diff_z, flag_diff_z)
		# Update the Pauli frame with the new corrections vis XORing
		pf[0] = pf[0] ^ pf_new_x ^ pf_flag_new_x
		pf[1] = pf[1] ^ pf_new_z ^ pf_flag_new_z

		# Set the last syndromes to the current ones
		last_syndrome_x = last_syndrome_x
		last_syndrome_z = last_syndrome_z

	return state, last_syndrome_x, last_syndrome_z, pf
\end{lstlisting}
\end{minipage}%
\end{figure*}%

\begin{figure*}
\begin{minipage}[t]{\textwidth}
\begin{lstlisting}[label={code:meas-rot}, caption={Rotate measurement to the correct logical basis},style=pseudo,]
def rotate_meas_basis(state: QuantumState, meas_basis: str) -> QuantumState:
	"""
	Rotate logical measurement basis.

	Args:
		state: The logical state.
		meas_basis: The logical measurement basis to measure in the end.

	Returns:
		A logical state.
	"""
	if meas_basis == 'Z':
		pass  # do nothing. we measure in Z by default

	elif meas_basis = 'X':
		state = logical_h(state)  # %\textcolor{MediumGray}{$H$}% on all data qubits

	elif meas_basis = 'Y':
		state = logical_sdg(state)  # %\textcolor{MediumGray}{$S$}% on all data qubits
		state = logical_h(state)  # %\textcolor{MediumGray}{$H$}% on all data qubits

	return state
\end{lstlisting}
\end{minipage}%
\end{figure*}%

\begin{figure*}
\begin{minipage}[t]{\textwidth}
\begin{lstlisting}[label={code:meas}, caption={A measurement in the logical Z basis.},style=pseudo,]
def logical_meas(state: QuantumState, meas_basis: str, last_syndrome_x: List[int, int, int], last_syndrome_z: List[int, int, int], pf: List[int, int]) -> int:
	"""
	Destructive logical measurement.

	Args:
		state: The logical state to measure.
		meas_basis: The logical measurement basis being measured in.
		last_syndrome_x: The last X-type syndromes measured (not including flagged ones)
		last_syndrome_z: The last Z-type syndromes measured (not including flagged ones)

	Returns:
		Logical measurement outcome.
	"""
	# Note: The logical measurement basis might have rotated outside of this function.

	# Measure each data qubit with a single-qubit Z-basis measurement.
	# Measurement labeling corresponds to the labeling of data qubits in Fig.%\textcolor{MediumGray}{~\ref{FlowChart}}%
	m1, m2, m3, m4, m5, m6, m7 = meas_z_data(state)  # get measurement results

	# Due to symmetry we can treat the X, Y, and Z basis in a similar manner:

	# Get logical measurement output of logical X, Y, or Z (as appropriate)
	meas_output = m5 ^ m6 ^ m7  # XOR measurements of logical operator on qubits 5, 6, 7

	# Get syndromes by XORing measurement bits together
	# Note: Due to the meas. basis, these results may be for X, Y, or Z type stabilizers
	s1 = m1 ^ m2 ^ m3 ^ m4
	s2 = m2 ^ m3 ^ m5 ^ m6
	s3 = m3 ^ m4 ^ m6 ^ m7
	syndromes = [s1, s2, s3]

	# Get change of syndrome depending on measurement basis
	if meas_basis == 'X':
		syndrome_diff = bitwise_xor(syndromes, last_syndrome_x)
	elif meas_basis == 'Y':
		syndrome_diff = bitwise_xor(syndromes, last_syndrome_x)
		syndrome_diff = bitwise_xor(syndromes, last_syndrome_z)
	elif meas_basis == 'Z':
		syndrome_diff = bitwise_xor(syndromes, last_syndrome_z)
	else:
		raise Exception()  # No other basis expected

	# Get correction from the syndrome diff as determined from measurement
	final_correction = decoder_2d(syndrome_diff)

	# Update logical measurement outcome based on the measured syndromes
	meas_output = meas_output ^ final_correction  # XOR together => flip outcome or not

	# Apply correction from Pauli frame to measurement outcome
	if meas_basis == 'X':
		meas_output = meas_output ^ pf[1]  # Apply logical Z
	elif meas_basis == 'Y':
		meas_output = meas_output ^ pf[0]  # Apply logical X
		meas_output = meas_output ^ pf[1]  # Apply logical Z
	elif meas_basis == 'Z':
		meas_output = meas_output ^ pf[0]  # Apply logical X
	else:
		raise Exception()  # No other basis expected

	return meas_output
\end{lstlisting}
\end{minipage}%
\end{figure*}%

\begin{figure*}
\begin{minipage}[t]{\textwidth}
\begin{lstlisting}[label={code:exp-result}, caption={Determining if the result is expected.},style=pseudo,]
def expected_result(meas_output: int, init_state: str, meas_basis: str) -> int:
	"""
	Determining whether a measurement result is consistant with expectation.

	Args:
		meas_output: Logical measurement output after correction. 0 => +1, 1 => -1
		init_state: The logical state intended to be initialized.
		meas_basis: The logical measurement basis measured in.

	Returns:
		A logical state.
	"""
	# Measurement basis and state pairs that should have 0/+1 outcomes
	zero_outcomes = {('X', '|+>'), ('Y', '|+i>'), ('Z', '|0>'),
		('X', '|T>'), ('Y', '|T>')}
	# Note, the measurements of the |T> state is not a measurement of fidelity but instead understood as the probability of being in the +1 eigenstate of the measured basis.

	# Measurement basis and state pairs that should have 1/-1 outcomes
	one_outcomes = {('X', '|->'), ('Y', '|-i>'), ('Z', '|1>')}

	if (meas_basis, init_state) in zero_outcomes:
		expected_val = 0

	elif (meas_basis, init_state) in one_outcomes:
		expected_val = 0

	else:
		# No other measurement basis and state pairs are expected to be evaluated.
		raise Exception('Unexpect init_state and measurement basis combination!')


	if expected_val == meas_out:
		success = 1
	else:
		success = 0

	return success
\end{lstlisting}
\end{minipage}%
\end{figure*}%

\begin{figure*}
	\begin{minipage}[t]{\textwidth}
		\begin{lstlisting}[label=code:decoder-2d, caption={The basic 2D decoder that infers if a logical error has occured.},style=pseudo,]
			def decoder_2d(syndrome_diff: List[int, int, int]) -> int:
				"""
				A 2D decoder that determines logical corrections. Due to symmetry this, can be used to decode stabilizer measurements of any Pauli type.

				Args:
					syndrome_diff: The change in syndromes compared to the last time they where measured.

				Returns:
					A bit representing whether a logical error has occurred.
				"""
				# Syndromes that indicate tha a single Pauli fault has flipped the logical operator representative on data qubits 5, 6, and 7.
				bad_syndromes = {[0, 1, 0], [0, 1, 1], [0, 0, 1]}

				if syndrome_diff in bad_syndromes:
					logical_error = 1
				else:
					logical_error = 0

				return logical_error
		\end{lstlisting}
	\end{minipage}%
\end{figure*}

\begin{figure*}
	\begin{minipage}[t]{\textwidth}
		\begin{lstlisting}[label=code:decoder-fix, caption={An additional decoder that updates the correction based on flag information.},style=pseudo,]
			def decoder_flag_update(syndrome_diff: List[int, int, int], flag_diff: List[int, int, int]) -> int:
				"""
				A 2D decoder that determines logical corrections. Due to symmetry this, can be used to decode stabilizer measurements of any Pauli type.

				Args:
					syndrome_diff: The change in syndromes compared to the last time they were measured.
					flag_diff: The change in flags compared to the last time syndromes were measured.

				Returns:
					A bit representing whether the 2D decoder's correction should be changed due to the relationship between flags and syndromes.
				"""
				# The following indicate hook faults have occured:

				# flag -> syndrome: 0 -> 1
				if flag_diff == [1, 0, 0] and syndrome_diff == [0, 1, 0]:
					change_correction = 1

				# flag -> syndrome: 0 -> 2
				elif flag_diff == [1, 0, 0] and syndrome_diff == [0, 0, 1]:
					change_correction = 1

				# flag -> syndrome: 1,2 -> 2
				elif flag_diff == [0, 1, 1] and syndrome_diff == [0, 0, 1]:
					change_correction = 1

				else:
					change_correction = 0

				return change_correction
		\end{lstlisting}
	\end{minipage}%
\end{figure*}%

\end{document}